\documentclass[12pt,preprint]{aastex}

\shorttitle{From Prestellar to Protostellar Cores}
\shortauthors{Aikawa et al.}

\begin{document}


\title{{\bf From Prestellar to Protostellar Cores II.  Time Dependence and Deuterium Fractionation} }


\author{Y. Aikawa}
\affil{Department of Earth and Planetary Sciences, Kobe University,
    657-8501, Kobe, Japan}
\email{aikawa@kobe-u.ac.jp}

\author{V. Wakelam, F. Hersant}
\affil{Univ. Bordeaux, LAB, UMR 5804, F-33270, Floirac, France\\
CNRS, LAB, UMR 5804, F-33270, Floirac, France}
%\affil{Universit$\acute{\rm e}$ Bordeaux, Observatoire Aquitain des Sciences de l'Univers,
%       2 rue de l'Observatoire, BP 89, F-33271 Floirac Cedex, France\\
%       CNRS, UMR 5804, Laboratoire dfAstrophysique de Bordeaux, 2 rue de l'Observatoire, 
%       BP 89, F-33271 Floirac Cedex, France}

\author{R.T. Garrod}
\affil{Department of Astronomy, Cornell University, Ithaca, NY 14853, USA}

\and

\author{E. Herbst}
\affil{Departments of Chemistry, Astronomy, and Physics, University of Virginia,
Charlottesville, VA 22904, USA}



\begin{abstract}
We investigate the molecular evolution and D/H abundance ratios that develop as star formation proceeds from a dense-cloud core to a protostellar core, by solving a gas-grain reaction network applied to a 1-D radiative hydrodynamic model with infalling fluid parcels. Spatial distributions of gas and ice-mantle species are calculated at the first-core stage, and at times after the birth of a protostar. Gas-phase methanol and methane are more abundant than CO at radii $r\lesssim 100$ AU in the first-core stage, but gradually decrease with time, while abundances of larger organic species increase.
The warm-up phase, when complex organic molecules are efficiently formed, is longer-lived for those fluid parcels in-falling at later stages. The formation of unsaturated carbon chains (warm carbon-chain chemistry) is also more effective in later stages; C$^+$, which reacts with CH$_4$ to form carbon chains, increases in abundance as the envelope density decreases. The large organic molecules and carbon chains are strongly deuterated, mainly due to high D/H ratios in the parent molecules, determined in the cold phase.
We also extend our model to simulate simply the chemistry in circumstellar disks, by suspending the 1-D infall of a fluid parcel at constant disk radii. The species CH$_3$OCH$_3$ and HCOOCH$_3$ increase in abundance in $10^4-10^5$ yr at the fixed warm temperature; both also have high D/H ratios.
\end{abstract}





\section{Introduction}

In star-forming cores, temperature and density vary both temporally and spatially.
The rates of chemical processes, which determine the molecular abundances in these cores, also
change according to the physical conditions. In prestellar cores with high densities and
low temperatures, various atoms and molecules are depleted onto grains, where they are hydrogenated
by grain-surface reactions. In the later, protostellar phase, on the other hand, molecules are desorbed
from the grains back into the gas phase in the central high-temperature regions. The molecular abundances in
the cores thus change dramatically in the course of star formation. 

It is well established that molecular clouds are generally not in chemical equilibrium;
the chemical timescale is comparable or longer than the dynamical timescale of the clouds.
For example, the adsorption timescale of gaseous species onto grain surfaces in cold
prestellar cores is
\begin{equation}
\left[\pi a^2 \sqrt\frac{8kT}{\pi m} S n_{\rm grain}\right]^{-1} \sim
1\times 10^6 \left(\frac{10^4 {\rm cm}^{-3}}{n_{\rm H}}\right)
\left(\frac{10^{-12}}{n_{\rm grain}/n_{\rm H}}\right)
\left(\frac{10^{-5} {\rm cm}}{a}\right)^2
\left(\frac{1.0}{S}\right){\rm [yr]},
\end{equation}
where $a$ and $n_{\rm grain}$ are the radius and number density of the grain particles, respectively,
and $S$ is the sticking probability \citep[e.g.][]{herbst93}. At the typical density of molecular clouds,
$n_{\rm H}\sim 10^4$ cm$^{-3}$, this timescale is comparable to the free-fall timescale of a cold core \citep[e.g.][]{spitzer73}
\begin{equation}
t_{\rm ff}=\sqrt{\frac{3\pi}{32G \rho}}\sim 4.3 \times
10^5 \left(\frac{10^4 {\rm cm}^{-3}}{n_{\rm H}}\right)^{1/2} {\rm [yr]}.
\end{equation}
Thus, hydrodynamic-chemical models are required to investigate theoretically the abundances and spatial distributions of molecules in star-forming cores.

Since the hydro-chemical models combine hydrodynamics and chemical reaction-network models,
there is a variety of previous work related to this topic. The chemistry in prestellar cores has been investigated using 
Bonner-Ebert spheres and/or the isothermal collapse model \citep[e.g.][]{aikawa01,keto10}.
The chemistry in protostellar cores is
often investigated using a pseudo-time-dependent gas-phase chemistry model with a constant temperature
($\sim 200$ K) and density ($n_{\rm H}\sim 10^6$ cm$^{-3}$), in which a high abundance of ice sublimates
is given as an initial condition \citep[e.g.][]{charnley92}.
While such simple models give chemical timescales of various species,
which is useful in the interpretation of protostellar cores, ice abundances should ideally be determined
by the gas-grain chemical model in the prestellar phase.

\citet{rodgers03} investigated the chemistry in a protostellar core using the inside-out collapse model \citep{shu77};
they calculated radial distributions of
molecular abundances at $10^2, 10^3, 10^4$ and $10^5$ yr after the collapse of the singular isothermal sphere.
Temporal variation of the temperature distribution, which is a critical parameter for the chemistry, was adopted from \citet{adams85}. The model shows the onion-skin type structure, in which
more volatile species are sublimated to the gas phase at outer radii, with the sublimation radius moving
outwards as the core becomes hotter. The model did not include a prestellar phase, since the inside-out collapse model corresponds to the main accretion phase, and the ice abundance was set as an initial condition.

\citet{lee04} went one step farther; they solved the chemistry in a core
starting from a prestellar core and evolving to a protostellar core self-consistently. Their physical model
of the core is a combination of a series of Bonner-Ebert spheres and inside-out collapse.
The spatial and temporal
variation of the temperature in the protostar is calculated by
radiation transfer, assuming the luminosity of the central star as a function of time. The model
clearly shows that the sublimation of ice (e.g. CO) affects the gas-phase chemistry significantly.
Their chemical model, however, does not include grain-surface reactions except for  H$_2$ formation;
dust-grain ice mantles are composed of molecules accreted directly from the gas phase.

\citet{garrod06}, on the other hand, constructed a comprehensive gas-grain chemical network with
various grain-surface reactions. Their chemical network is solved with simple physical models
in which the density is constant and the temperature increases with time. They showed that
some large organic species are efficiently formed on grain surfaces starting at temperatures of 
30-40 K and sublimated at higher temperatures. Their model accounts for the high abundance
of large organic molecules such as methyl formate (HCOOCH$_{3}$) and dimethyl ether (CH$_{3}$OCH$_{3}$) in low-mass protostellar cores, known as hot corinos
\citep{ceccarelli07}.
\citet{aikawa08}, hereinafter Paper I,  then applied the chemical network of \citet{garrod06} to a 1-D radiation-hydrodynamic
model of a star-forming core by \citet{masunaga00}. As expected,  organic species such as
CH$_3$CN and HCOOH are formed in the warm-up phase and become abundant in the central region
($T\gtrsim 100$ K) of the protostellar core. In addition, the model shows that (unsaturated) carbon-chain species
increase in abundance inwards at $T\sim 25$ K, which is observed as the so-called ``Warm Carbon Chain Chemistry (WCCC)" \citep{sakai08}.

In this work, we update and improve the model of Paper I.
The aim of the work is to investigate
\begin{enumerate}
\item{the spatial distribution of molecules at several evolutionary stages from a prestellar core to
a protostellar core,}
\item{the evolution of molecular D/H ratios,}
\item{how the chemical composition in a disk differs from that in cores.}
\end{enumerate}
Paper I showed spatial distributions of molecules only at $9.3\times 10^4$ yr after the protostellar birth,
although the model starts from a prestellar core. Here we show molecular distributions at assorted
evolutionary stages to see how the distributions and peak abundances of hot corino species and carbon
chains vary with time. We also extend the chemical reaction network to include both singly- and multiply-deuterated species.
It is well established that molecular D/H ratios are enhanced by exothermic exchange reactions
at low temperature. Once the star is formed, the core becomes warmer and the backward reactions
(endothermic-exchange reactions) become
efficient. How the D/H ratios vary after protostellar birth is of importance to link the observed
high D/H ratios in the ISM to the isotopic ratios observed in planetary matter such as comets and
meteorites. In order to investigate such a link, it is also of great interest  to determine how the chemical composition
in a disk differs from that of cores. For the material in an  infalling envelope, the duration of the
warm-up phase and warm chemistry is determined by the size of the warm region divided by the
free-fall velocity, which is rather short, as shown in Paper I. In circumstellar disks, on the other
hand, material can stay in warm regions for longer timescales, since the disk is supported by rotation,
at least partially. Because our model is spherical and thus does not include the structure and dynamics of disk formation,
we briefly investigate the effect of prolonged warm temperature chemistry in the disk by calculating
pseudo-time-dependent models with the initial abundances set by the infalling core model.

In the following section, we briefly describe our physical model of the star-forming core and the chemical
reaction network (\S 2). In \S 3, we report the spatial distribution of molecular abundances and D/H ratios
at assorted evolutionary stages. In \S 4, we discuss the effect on the chemistry of 2D structure in the core;
we calculate a pseudo-time-dependent model of a warm disk-like phase. We also compare our model results with
observational studies, and discuss the effect of  grain-surface abstraction and substitution reactions on D/H ratio
of methanol. We summarize our conclusions in \S 5.

%A focal problem today in the dynamics of globular clusters is core collapse.  
%It has been predicted by theory for decades \citep{hen61,lyn68,spi85}, but
%observation has been less alert to the phenomenon. For many years the
%central brightness peak in M15 \citep{kin75,new78} seemed a unique anomaly.  
%Then \citet{aur82} suggested a central peak in \object{NGC 6397}, and a 
%limited photographic survey of ours \citep[Paper I]{djo84} found three more 
%cases, \objectname{NGC 6624}, \objectname[M 15]{NGC 7078}, and 
%\object[Cl 1938-341]{Terzan 8}), whose sharp center had often been 
%remarked on \citep{can78}.  

\section{Model}

\subsection{Physical model of a star-forming core}

\begin{figure}
%\epsscale{.80}
\plotone{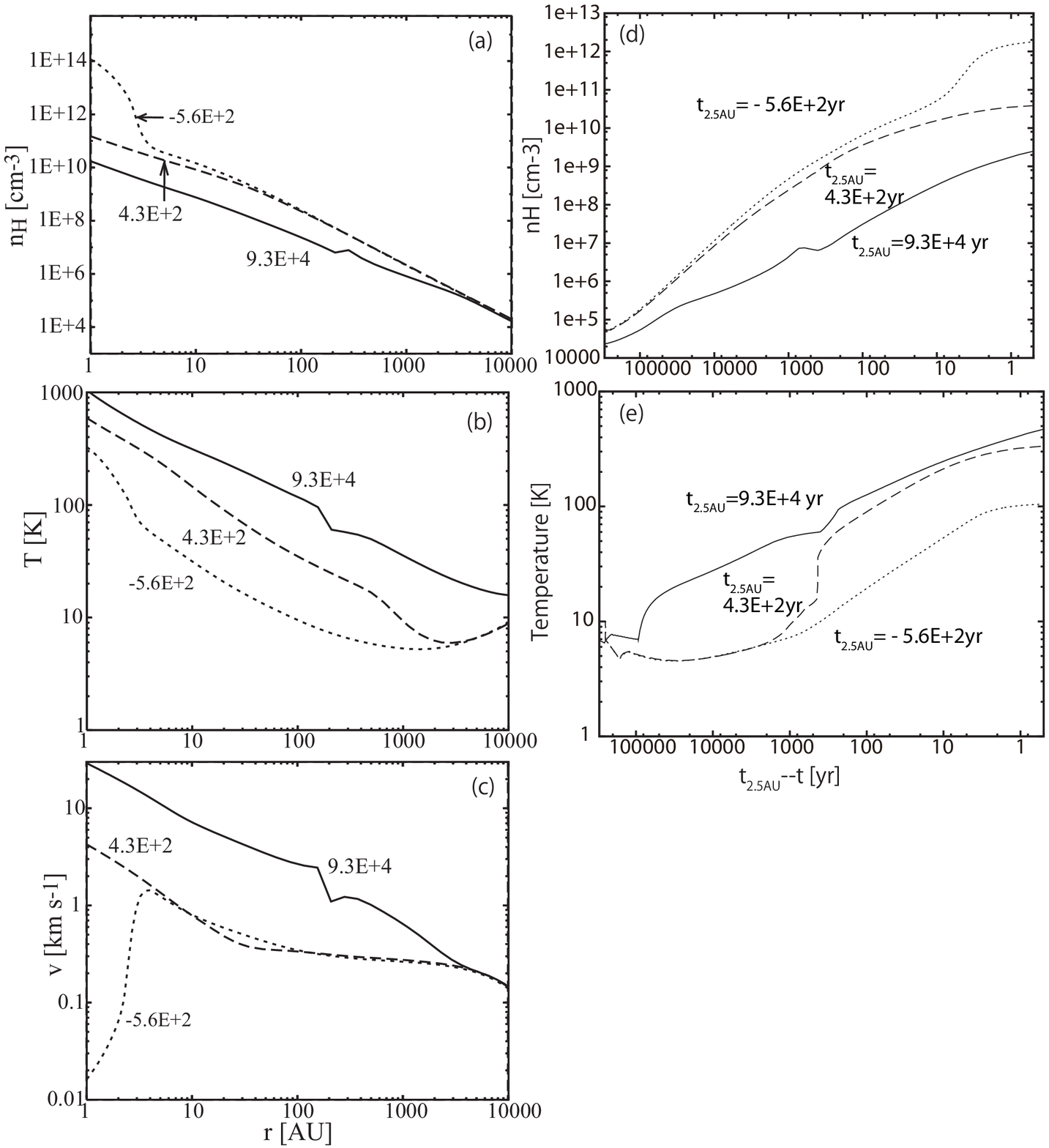}
\caption{($a-c$) Radial distribution of density ($n_{\rm H}$), temperature and infalling
velocity at $t=-5.6\times 10^2$ yr ({\it dotted line}) , $4.3\times 10^2$ yr
({\it dashed line}), and $9.3\times 10^4$ yr ({\rm solid line}), where $t=0$ is defined
by the birth of a protostar. ($d-e$) Temporal variation of density and temperature
in fluid parcels which reach $r=2.5$ AU at $t=-5.6\times 10^2$ yr, $4.3\times 10^2$ yr,
and $9.3\times 10^4$ yr.\label{dist_phys}}
\end{figure}

As in Paper I, we adopt the 1D (spherical) radiation hydrodynamic model of  low-mass star formation
by \citet{masunaga98} and \citet{masunaga00}. 
Initially, the central density of the molecular cloud core is
$\rho=1.4 \times 10^{-19}$ g cm$^{-3}$, which corresponds to a number density of hydrogen nuclei
$n_{\rm H}\sim 6 \times 10^4$ cm$^{-3}$. The outer boundary is fixed
at $r=4 \times 10^4$ AU, so that the total mass is 3.852 $M_{\odot}$, which exceeds
the critical mass for gravitational instability. The contraction is almost isothermal
as long as the cooling rate overwhelms the compressional
heating, but eventually the latter dominates and the temperature rises
in the central region. Afterwards, the first core, which is a hydrostatic core of H$_2$ gas, is formed.
When the central temperature reaches $\sim 2000$ K (several hundred years after the first core formation),
H$_2$ starts to dissociate and the first core collapses.
The central region becomes hydrostatic again when the dissociation and ionization of hydrogen are completed;
this signifies the birth of a protostar.
In our model, the prestellar core evolves to the protostellar core in $2.5 \times 10^5$ yr.
After the birth of the protostar, the model further follows the
evolution for $9.3 \times 10^4$ yr, during which the protostar grows by mass accretion from
the envelope. At each evolutionary stage, the model gives the total luminosity of the core and the
radial distribution of density, temperature, and infall velocity at $r\gtrsim 10^{-4}$ AU self-consistently.
The original model by \citet{masunaga00} included wavelength-dependent radiation transfer involving
the opacities of the gas \citep{iglesias96,alexander94} and dust \citep{preibisch93} to obtain both the temperatures of the gas
and the various components of dust (silicate, carbon and ice) separately. For simplicity, we adopt their gas temperature
and assume that it is equal to the dust temperature, since the differences between the gas and
dust temperatures are not siginificant in the region we are interested in.
Figure \ref{dist_phys} ($a-c$) shows the distribution of density, temperature, and infall velocity
at assorted evolutionary stages, in which we present the molecular distribution in
\S 3: $t=-5.6\times 10^2$ yr, $4.3\times 10^2$ yr and $9.3\times 10^4$ yr.
{\em Here we define $t=0$ as the moment of the birth of a protostar.}
The total luminosity of the core is about $5.9\times 10^{-3} L_{\odot}$ at $t=-5.6\times 10^2$ yr, $0.88 L_{\odot}$
at $t=4.3\times 10^2$ yr, and $24 L_{\odot}$ at $9.3\times 10^4$ yr.
The first core
is formed at around $t= -5.6\times 10^2$ yr; Figure \ref{dist_phys} ($a-c$) shows that at this time a hydrostatic core
of a few AU size is formed. It should be noted that the density in the envelope decreases with time
after the protostar is formed, while the density increases with time in the prestellar stage.
More detailed explanations of core evolution can be found in \citet{masunaga98}, \citet{masunaga00}
and Paper I.

Figure \ref{dist_phys} ($d-e$) shows the temporal variation of density and temperature in  fluid parcels
that fall to $r=2.5$ AU at $t=-5.6\times 10^2$ yr, $4.3\times 10^2$ yr and $9.3\times 10^4$ yr.
Naturally, the density and temperature increase with time.
As can be seen in Eq. (2), the free-fall timescale is proportional to
$\rho^{-1/2}$. While the core temperature rises
in the central region, the infall velocity increases toward the center until the fluid parcel
hits the surface of the central hydrostatic core (the first core or the second core). Hence the temporal variation of
density and temperature accelerates in the infalling fluid parcels. In order to highlight the rapid rise near and at
the final stage, the horizontal axis in Figure \ref{dist_phys} ($d-e$) is set to be the logarithm of
$t_{\rm 2.5AU}-t$, where $t_{\rm 2.5 AU}$ corresponds to $t=-5.6\times 10^2$ yr for the dotted line,
$4.3\times 10^2$ yr for the dashed line, and $9.3\times 10^4$ yr for the solid line.

The temporal variation of temperature is critically important for chemistry. Garrod \& Herbst (2006)
show that the grain-surface reactions of heavy-element radical species are efficient in a 
temperature range of $\approx 30 - 100$ K. Once the temperature gets higher than $\sim 100$ K, most
volatile species are sublimated to the gas phase, where they begin to undergo gas-phase reactions.
In Figure \ref{dist_phys} ($d-e$), we can easily read the timescale of these warm phases. For example,
the fluid parcel that reaches $r=2.5$ AU at $t=9.3\times 10^4$ yr stays at a temperature of $30-100$ K
for $\sim 10^4$ yr.  After the fluid parcel gets warmer than 100 K, it falls to
the central star in a few 100 yr. The fluid parcels that reach $r=2.5$ AU at earlier
time spend even less time in the warm regions.

In the original model by \citet{masunaga00}, the core starts contraction immediately, since the core
mass exceeds the critical mass for gravitational instability. For our chemistry model, however, we
need to set up the initial molecular abundances of this dense molecular core by assuming a simple history of
fluid parcels before collapse. Since the formation of molecular cloud cores is out of the scope of this work,
we simply assume that the core keeps its hydrostatic structure for $1\times 10^6$ yr, implicitly assuming that
turbulence supports it. In addition, we assume that the temperature in the pre-collapse phase is 10 K, as
typically observed
in molecular clouds. After $1\times 10^6$ yr, we lower the temperature smoothly but rapidly to the initial value
in \citet{masunaga00}, $\sim 7$ K.

\subsection{Chemical reaction network}

We solve the rate equations of the gas-grain reaction network of \citet{garrod06}
under the time-dependent physical conditions of each fluid parcel, to obtain
the radial distributions of molecular abundances both in the gas phase and ice mantles at
assorted evolutionary stages. We use the gas-grain code Nautilus
\citep{hersant09, semenov10} to solve the rate equations. A layered structure of the ice mantle is not
considered (see \S 3.2).
The chemical reaction network model and parameters are essentially the same as in paper I,
except for the following three updates:
(i) Some gas-phase reaction rates are updated based on the KIDA database (http://kida.obs.u-bordeaux1.fr/).
(ii) We included reactions between carbon-chain species with H atoms or H$_2$ molecules, based on
\citet{harada10}. \citet{hassel11} showed that these reactions significantly change the evolution
of radical carbon-chain species at temperatures $\gtrsim 100$ K. We do not, however, discriminate between
isomers such as c-C$_3$H$_2$ and H$_{2}$CCC, which was done by \citet{hassel11}.
(iii) We include multi-deuterated species according to the following procedures.

First, the species list of the network model is extended to include mono-, doubly-, and triply-deuterated
species. Then the reaction network, both in the gas phase and on grain surfaces, is extended to
include these deuterated species. For example, if we have the reaction AH$^+$ + BH $\rightarrow$
A + BH$_2^+$ in the original reaction list, we add\\
AD$^+$ + BH $\rightarrow$ A + BHD$^+$ \\
AH$^+$ + BD $\rightarrow$ A + BHD$^+$ \\
AD$^+$ + BD $\rightarrow$ A + BD$_2^+$ \\
assuming that the rate coefficients are the same as for the original reaction. If there is more than
one set of possible products,\\
AHD + BD $\rightarrow$ ABH + D$_2$ \\
AHD + BD $\rightarrow$ ABD + HD, \\
for example, we set the branching ratio statistically \citep[e.g.][]{furuya11}.
In the above example, the branching ratio for the former and
latter product channels is 1:2.
We also assumed statistical branching ratios in dissociative recombination reactions, except for
the recombination of deuterated H$_3^+$; this assumption can result in 
underestimates of
the molecular D/H ratio by a factor of two or so, since channels with a hydrogen atom as a product
are more likely than analogous ones with a D atom due to the greater speed of the H atom \citep{lepetit03}. 
For the recombination of deuterated H$_3^+$, we adopted branching ratios listed in \citet{roberts04}.
Indeed, these branching ratios are higher for reactions producing an H atom than the statistical value.
For simplicity, we do not follow the position of the deuterium atoms in the reactions which include deuterated isomers.
For example, in the reaction of  H$_2$CO + OD, the branching ratio of the three product channel \\
H$_2$CO + OD $\rightarrow$ HCOOH + D \\
H$_2$CO + OD $\rightarrow$ DCOOH + H \\
H$_2$CO + OD $\rightarrow$ HCOOD + H \\
is 1:1:1 in our current model, although the last branch would be the most probable in reality. So we do not aim to predict the
abundance ratios of deuterated isomers.
%One exception is methanol, for which we assume that a direct exchange reaction of
%CH$_3$OH + D makes only  CH$_2$DOH, not CH$_3$OD, referring to the laboratory experiment by \citet{nagaoka05}
%(see \S 5.2).

Finally we included reactions listed in \citet{millar89} and
\citet{roberts04}; these are mostly exchange reactions that cause the initial isotopic fractionation.
At low temperatures, of several tens of Kelvin or less, deuterium fractionation is mainly triggered by H$_3^+$ + HD $\rightarrow$
H$_2$D$^+$ + H$_2$ (and its multi-deuterated analogues), which is exothermic by 230 K.
%The efficiency of the backward reaction depends on if the H$_2$ is in ortho- or para state, since the internal energy
%of o-H$_2$ helps overcome the endothermicity. In the present work, we assume that all H$_2$ is in para state.
The effective rate of the backward reaction depends on the ortho/para ratio of H$_2$, since the internal energy of ortho H$_2$
helps overcome the endothermicity. But the ortho/para ratio of H$_2$ is a complicated issue \citep[e.g.][]{flower06},
and there is so far no observational evidence for a high abundance of ortho H$_2$ in molecular clouds.
Recently, \citet{watanabe10} measured the spin temperature of H$_2$ formed on ice in the laboratory, obtaining a value greater than approximately 200 K. But they also found that ortho H$_2$ that is trapped on the ice surface
is converted to para H$_2$ after formation. Such conversion of ortho to para has never been considered in the
prediction of the o/p ratio of H$_2$. In this work, we do not consider the o/p ratio; i.e. we assume the backward reaction
of H$_3^+$ + HD is endothermic by 230 K.

Inclusion of multi-deuterated species naturally increases the number of species and reactions.
In total, our model includes 1564 species and 45371 reactions, while the original model includes
655 species and 6309 reactions.

As elemental abundances, we adopt the so-called ``low-metal" values (Table 1 of \citet{aikawa01}).
The species are assumed to be initially in the form of atoms or atomic ions except for hydrogen, which is
entirely in its molecular form. The elemental D/H ratio is set to $1.5\times 10^{-5}$ \citep{piskunov97}. 
All deuterium resides initially in the HD molecule. The cosmic-ray ionization rate for H$_{2}$ is
set to $1.3 \times 10^{-17}$ s$^{-1}$. 

The sticking probability for gaseous
neutral species onto grains is set to 0.5, independent of temperature. Although
the probability may be lower at high temperatures ($T\gtrsim$ 100 K),
the value of the sticking probability becomes unimportant once evaporation becomes dominant.
We assume the same set of adsorption energies ($E_{\rm ads}$) as \citet{garrod06}, corresponding to a
grain surface covered by water ice. The adsorption energies
for deuterated species are set to the values used for their normal isotopes, with the exception of D atoms, whose adsorption energy is set 21 K higher (471 K) than that of
H atoms (450 K), following \citet{caselli02}.
The grain-surface species can desorb to the gas phase via thermal evaporation and two non-thermal
processes: cosmic-ray desorption \citep{hase93} and desorption via exothermic
association reactions on grain surfaces \citep{williams68,garrod06,garrod07}.
Recent laboratory work and molecular dynamics simulations indicate that photodesorption could also
be important \citep{andersson08,oberg09a,oberg09b,munozcaro10,arasa11}.
There are several possible mechanisms for photodesorption:
Molecules are dissociated into radicals and atoms, which then desorb separately to the gas phase; the dissociation products may first recombine on the grain and then desorb \citep{andersson08}; or the dissociation products may kick out neighboring ice-mantle species.
Our model includes the second of these mechanisms, through the explicit inclusion of two separate processes; the photodissociation into radicals of grain-surface species by interstellar
and cosmic ray-induced UV photons (see below), and the partial evaporation of the products of exothermic surface reactions \citep{garrod07}.

We assume the Langmuir-Hinshelwood mechanism for grain surface reactions;
species can diffuse on grain surfaces by thermal hopping and react with each other
when they meet. No quantum tunneling was assumed in the migration, even for H and D atoms \citep{garrod07}.
The barrier for migration is set at 50\% of the adsorption energy.  We adopt the modified-rate approach for grain-surface reactions that include H and D atoms \citep{stantcheva01,caselli02}.
In paper I, we did not adopt the modified-rate approach, since its significance is 
reduced in the model when assuming only thermal hopping of H atoms on the grain surface,
compared with a model that includes quantum tunneling \citep{ruffle00}.
In the present work, however, the grain-surface reactions with D atoms could still be
accretion-limited rather than migration-limited, and thus the modified rates can be important.

Extinction of interstellar UV radiation is calculated from the column density of hydrogen nuclei
($N_{\rm H}$) from the core outer edge to each shell by the formulation
$A_{\rm V} = N_{\rm H}/(1.59 \times 10^{21} {\rm cm}^2)$ mag.
Initially, the visual extinction from the outer edge ($r= 4\times 10^4$ AU) to the core
center is about 5.5 mag, and the outermost shells in which we calculate chemistry are located at
$1\times 10^4$ AU, where $A_{\rm V}\sim 1$ mag. Assuming that our model core is embedded
in ambient clouds, we add 3 mag to the visual extinction obtained above, and ignore
photodissociation of CO and H$_2$, which should be self- and mutually-shielded from the interstellar UV
at $A_{\rm v}\gtrsim 1$ mag \citep[e.g.][]{vandishoeck88}.
We also neglect the photodissociation of HD and D$_2$, since it is not clear if and how much they are shielded by H$_2$. This assumption would not much affect our results
for molecular D/H ratios, since the interstellar UV is extinguished by $\ge 3$ mag.
The rates of photodissociation and photoionization by cosmic ray-induced UV radiation, on the other hand,
are roughly independent of $A_{\rm v}$ and proportional to the cosmic-ray ionization rate, although the coefficients
vary among molecular species. The photodissociation rates of HD and D$_2$ by cosmic-ray induced UV are set to be
the same as for H$_2$.
UV radiation from the central protostar is not considered in our model; since the model is 1D,
UV radiation is absorbed in close vicinity to the source.

\section{Results}
\subsection{Temporal Variation of Molecular Distributions}

Figure \ref{dist_abun} ($a-b$) shows the radial distribution of simple molecules at $t=-5.6\times 10^2$ yr,
$4.3\times10^2$ yr, and $9.3\times 10^4$ yr.  These simple molecules comprise four major neutrals: 
CO, NH$_{3}$, H$_2$O and N$_2$. Major ions and electron abundance (i.e., the fractional ionization) are also plotted.
A simple view of the radial distribution of major species is as follows: as the radius gets smaller, and the density
higher, the species first tend to be depleted from the gas onto the ice. Finally, as the temperature rises, sublimation
occurs and the species return rapidly and sharply to the gas at the so-called sublimation radius. Little change is seen in
gas-phase values at yet smaller radii, due to the rapid collapse. Such a radial distribution of gas-phase
abundance is consistent with the ``drop abundance profile" derived from multi-line observations of CO
and H$_2$O toward protostars \citep[e.g.][]{jorgensen04, cousten12}.
The sublimation radii of CO, NH$_3$, H$_2$O and N$_2$ increase with time, as
the core temperature rises, so that the gas-phase material extends farther from the core.
The CO depletion factor is defined as the ratio of the canonical value of CO, which is close to the gaseous
elemental abundance of carbon ($8\times 10^{-5}$ in our model), to the actual abundance of gaseous CO. The depletion factor
increases inward as the density increases, until it suddenly decreases at the sublimation radius.
The peak value of the CO depletion factor decreases with time, because the density decreases. We note that the CO depletion
factor does not reach unity even inside the sublimation radius in our model; i.e. CO abundance inside the sublimation radius
is smaller than the canonical value, especially at early stages, because CH$_3$OH and CH$_4$ are slightly more abundant
than CO. Low CO abundances inside the sublimation radius have recently been observed toward several protostars \citep{alonso10,yildiz12,fuente12}.
%We plot the abundance of electrons (i.e., the fractional ionization) and major cations in Figure \ref{dist_abun} ($a-c$). 

The fractional ionization decreases toward the center,
because it is generally proportional to $n_{\rm H}^{-1/2}$. 
It is noteworthy that HCO$^+$, which is
often used as a radio tracer of dense gas, decreases as two species with a higher proton affinity than CO (H$_2$CO and
NH$_3$) sublime into the gas phase. 

The distribution of complex organic species is shown in Figure \ref{dist_abun} ($c$).
At $t=-5.6\times 10^2$ yr, the temperature is lower than $\sim 10$ K at $\gtrsim 100$ AU (Figure \ref{dist_phys}).
In these cold regions, CH$_3$OH ice is as abundant as $1.62\times 10^{-5}$, and HCOOH ice and CH$_3$CN ice
are $\sim 10^{-9}$ relative to  hydrogen nuclei. CH$_3$OH ice is formed by the hydrogenation of CO ice, while
HCOOH is mainly formed in the gas phase via the dissociative recombination of CH$_3$O$_2^+$,
which is a product of the radiative-association reaction HCO$^+$ + H$_2$O. A fraction of the gas-phase HCOOH is adsorbed onto grains
before subsequent gas-phase reactions. CH$_3$CN is formed by association reactions (which add C and H
to CN) on grain surfaces. Because these species are formed by low-temperature chemistry, the abundances at this stage
 depend on the duration of the pre-collapse model assumed.
HCOOCH$_3$ ice, on the other hand, increases inwards at 10 AU; it is formed at warm temperatures, starting at 30-40 K.

At $t=4.3\times 10^2$ yr, the warm region of $T\gtrsim$ 100 K expands to $\sim 10$ AU, where complex
organic species  sublime. 
%HCOOCH$_3$ is increased significantly compared with $t=-5.6\times 10^2$ yr.
At $t=9.3\times 10^4$ yr, many organic species are formed via grain surface reactions
and gas phase reactions of sublimated species at $T\gtrsim$ 30 K (\citet{garrod06}; Paper I).
Our model shows that the abundance of large organic species increases with time at the central region
not only because the dust temperature exceeds  their sublimation temperatures, but also because the duration
of the warm chemistry is longer for fluid parcels in later stages than those in earlier stages
(see Figure \ref{dist_phys}). It should also be noted that the abundance rise is not homogeneous;
relative abundances among complex organic species (e.g. CH$_3$CHO/HCOOCH$_3$) vary with time.

Figure \ref{dist_abun}($d$) shows the distribution of carbon-chain species. 
At $t=9.3\times 10^4$ yr, carbon chains increase inwards at the sublimation radius
of CH$_4$ ($T\sim 25$ K, $r\sim 2000$ AU), showing that the chemistry is indeed described by WCCC; the CH$_4$ reacts with C$^+$ to form
C$_2$H$_3^+$, and unsaturated hydrocarbons are formed via subsequent gas-phase reactions and
a partial contribution of grain surface hydrogenation (\citet{sakai08};Paper I; \citet{hassel08}).
At earlier stages ($t=-5.6\times 10^2$ yr and $4.3\times 10^2$ yr), the formation of carbon-chains at the
CH$_4$ sublimation radius is much less significant. Although the CH$_4$ abundance is as high as
$2.5\times 10^{-5}$, the other reactant to start WCCC, C$^+$, is deficient.  Specifically, while the abundance of
C$^+$ at the sublimation radius of CH$_4$ is as high as $\sim 1\times 10^{-10}$ at $t=9.3\times 10^4$ yr,
it is $4\times 10^{-15}$ at $t=-5.6\times 10^2$ yr, because of the high density. Although some carbon chains
are formed at the CH$_4$ sublimation zone at $t=4.3\times 10^2$ yr, the increment is much smaller than that at
$t=9.3 \times 10^4$ yr.
%they are depleted onto grains within a short timescale because of
%their high adsorption energies compared with CH$_4$ and the high density of the region.
The abundances of carbon chains (e.g. C$_3$H$_2$) at the central region of these early phases are thus determined
mostly by the sublimation of icy carbon chains formed in the cold era.

It should also be noted that C$_4$H and C$_2$H are depleted at radii less than a few hundred AU at $t=9.3\times 10^4$ yr,
which is consistent with observations of L1527 \citep{nami10}.
In these hot-temperature regions they react with H$_2$ and are converted to C$_4$H$_2$ and C$_2$H$_2$, respectively
\citep{hassel11}. C$_3$H$_2$, on the other hand, does not deplete at the central region, while \citet{nami10}
found that the cyclic form, c-C$_3$H$_2$, does indeed exhibit a central dip. \citet{hassel11}, who distinguish c-C$_3$H$_2$ from H$_{2}$CCC,
also show that c-C$_3$H$_2$ survives in hot regions ($\sim 100-200$~K) for several $10^5$ yr, while the carbene form
is destroyed via reaction with H$_2$ over a shorter time scale at $T\gtrsim 140$ K.
Additional destruction paths for c-C$_3$H$_2$ are therefore required to account for the observations.

%[SATURATED CnHm should be plotted? Their contribution to WCCC should also be discussed here?]

\subsection{Molecular D/H ratios}

Figure \ref{dist_DH} shows the spatial distribution of selected mono-deuterated species, as well as D$_3^+$,
and their normal isotope counterparts at $t=-5.6\times 10^2$ yr, $4.3\times 10^2$ yr and $9.3\times 10^4$ yr.
The D/H ratios of assorted molecules are shown in the bottom panels in Figure \ref{dist_DH}.
%Although our network model includes multiply deuterated species,
%we show mainly mono-deuterated species so that the plot is not too busy.
In low-temperature chemistry ($T\sim 10$~K), gas-phase molecules are enriched with deuterium due to exothermic exchange
reactions such as H$_3^+$ + HD $\rightarrow$ H$_2$D$^+$ + H$_2$ (and analogous reactions for multiply deuterated
species; Millar, Bennett \& Herbst 1989; Roberts, Herbst \& Millar 2004). The backward reactions, e.g. H$_2$D$^+$ + H$_2$ $\rightarrow$ H$_3^+$ + HD, are 
endothermic (typically by several 100 K) and thus are negligible at low temperature ($\sim 10$ K).
The D/H ratio for H$_{3}^{+}$ is not governed by these two reactions alone, as H$_3^+$ and its deuterated isotopomers
are destroyed by reactions with CO and electrons.
In the zone where gaseous CO is depleted, however, the D/H ratio for H$_3^+$ 
is further enhanced.  For example, at $t=-5.6\times 10^2$ yr, D$_3^+$
becomes the major cation in the CO depletion zone at radii of a few tens to several hundred AU (see Fig. \ref{dist_DH}). 
Although the exothermic exchange reactions
are limited to several species, the high D/H ratio propagates to other molecules
via gas-phase reactions; e.g. H$_2$D$^+$ gives its deuteron to other molecules and atoms via ion-molecule reactions. 
It is essential to include multi-deuterated species, such as D$_3^+$,
to calculate the molecular D/H ratio, since the multiply-deuterated species
can propagate D to other species more efficiently than can mono-deuterated ones.
The high D/H ratio also propagates to the grain surface; deuterated H$_3^+$ dissociatively
recombines with electrons to produce D atoms, which are
adsorbed onto grains and deuterate the ice-mantle species.

Once the core gets warmer than $\sim 20$ K, CO sublimation terminates the extreme deuteration (XD/XH$\gtrsim 1$)
of H$_3^+$. At the same time, the endothermic exchange reaction, H$_2$D$^+$ + H$_2$ 
$\rightarrow$ H$_3^+$ + HD, becomes effective and dominates in the destruction of deuterated H$_3^+$ at higher temperatures.
Since H$_2$ is very abundant, the ratio of H$_2$D$^+$ to H$_3^+$ rapidly reaches its equilibrium value, which is
$\sim 1.3\times 10^{-5} (T/300 {\rm K})^{-0.8} \exp(-230 {\rm K}/T)$,
i.e. $4\times 10^{-3}$ and $3\times 10^{-4}$ at 30 K and 50 K, respectively \citep{millar89}.
Hence H$_2$D$^+$ and HD$_2^+$ may be good observational temperature probes in low- to intermediate-temperature regions ($\lesssim 50$ K) of the core.
%Molecular ion CH$_3^+$
%and C$_2$H$_2^+$ can maintain relatively high D/H ratio $\gtrsim 10^{-3}$ upto $\sim 100$ K, because they under go
%exchange reactions with HD, which are more endothermic than H$_3^+$ + HD.%KOKOCHECK

Neutral species have high D/H ratios even in warm regions. These neutral species,
such as NH$_3$, CH$_4$, and CH$_3$OH, are formed more efficiently via grain-surface reactions than by gas-phase
reactions in the cold era. Inside the sublimation radii, the abundances of the sublimates overwhelm those of the
products of gas-phase reactions. The icy material is highly deuterated by surface reactions with D atoms
as described above, although the D/H ratios inside
the sublimation radii are lower than the ratios of their gaseous counterparts in the CO depletion zone, since
ices are formed and accumulated throughout the cold era, including the time before significant CO depletion occurs.
For the same reason, the D/H ratios of the sublimates (i.e. inside the sublimation radii) are higher in earlier
stages only by a factor of a few.
The D/H ratio of these gaseous neutral species is constant down to $r=2.5$ AU, since their destruction time scales
are longer than a free-fall time scale.

At $t=9.3\times 10^4$ yr, various complex organics are formed via both gas-phase and grain-surface reactions;
various carbon chains and large organics are formed via WCCC at $T>25$ K and via radical-radical surface reactions
at $T\gtrsim 30$ K, respectively.
Their D/H ratios are high, mainly because they are formed from molecules with high D/H ratios.
As an example, methyl formate (HCOOCH$_{3}$) can be formed by the surface radical-radical reaction between the radicals HCO 
and CH$_3$O. Since both HCO and CH$_3$O have high D/H ratios, DCOOCH$_{3}$ and HCOOCH$_2$D are also formed.
The HCOOCH$_2$D is three-times more abundant than DCOOCH$_3$, since we assume the statistical branching rather than
following the position of D in each reaction.

It is well known that the exchange reactions of CH$_3^+$ + HD and C$_2$H$_2^+$ + HD are more exothermic than
H$_3^+$ + HD, and thus can be effective even at relatively warm temperatures (several tens of K).
However, they do not much contribute to the deuteration of WCCC species
in our model, probably because carbon chains and CH$_4$ already have high D/H ratios when the fluid parcels enter
the warm region, and because the duration of the warm temperature (around 30 K) is short.
The D/H ratios of carbon-chain species change little, even if we neglect these two exchange reactions.

The exchange reaction of OH + D $\rightarrow$ OD + H, on the other hand, has an even higher exothermicity
($\Delta E/k \sim 810$ K), and significantly deuterates OH at $t=9.3\times 10^4$ yr at radii of 100 -- 1000 AU, where
HCOOH has a higher D/H ratio than other organic species due to the reaction of sublimated H$_2$CO with OD.

Radial distributions of assorted multiply-deuterated species are shown in Figure \ref{dist_multiD},
together with mono-deuterated and normal isotopologues. The abundance ratios of multi- to mono-deuterated species
are mainly determined statistically, e.g. XD$_2$/XHD $\sim$ XHD/XH$_2$. For CH$_3$OH and H$_2$CO, however,
the multi- to mono ratio is lower than the mono-deuterated to normal isotope ratio.
This effect could be due to the activation barriers
in their formation path (i.e. in the reactions H + CO and H + H$_2$CO);
the tunneling rate through the activation barrier is lower for heavier isotopomers (see discussion in \S 4.3).
For NH$_3$ and CH$_4$,
on the other hand, the multi-deuterated species are more abundant than expected from their mono-deuterated to
normal isotope ratio. Although they are mainly formed on grain surfaces, gas-phase formation also contributes.
In the gas-phase, multiply-deuterated NH$_3$ and CH$_4$ are efficiently formed in the CO depletion zone.
Since the abundances of multiply-deuterated species are generally low, the efficient formation in the CO depletion zone
effectively increases its abundance relative to the normal isotope.
At $t=9.3\times 10^4$ yr, multi-deuterated NH$_3$ and CH$_4$ are much less
abundant than in early stages, because the CO depletion is much less significant.

In our model, we do not consider the layered structure of ice mantles. In reality, ice mantles
have layered structures and the ices with the highest D/H ratios, which are formed at the latest stage of the cold
prestellar phase, would lie on the very surface layer of the ice mantle, and would be the first to be sublimated.
This effect creates radial gradients of D/H ratios. For example, the CH$_4$ formed in the CO depletion zone would
have a higher D/H ratio than that formed in the early molecular cloud stage. Since the CH$_4$ formed in the early stage would
coexist within H$_2$O ice, mainly the highly deuterated CH$_4$ (in the surface layers) would be sublimated
at $T\sim 25$ K; e.g. at $t=4.3\times 10^2$ yr the gaseous CH$_4$ abundance would be lower and CH$_3$D/CH$_4$
ratio would higher in $r=10-100$ AU than shown in Figure \ref{dist_DH}. Once the temperature reaches $T\gtrsim 150$ K,
however, the dominant ice component, H$_2$O ice, sublimates together with other species formed in the early stages of
molecular clouds. Then the abundance and D/H ratio of CH$_4$, for example, would be the same as shown in Figure \ref{dist_DH}
inside $\sim 10$ AU at $t=4.3\times 10^2$ yr.
%But such gradients should be limited to a region of $T\sim$
%$30 - 100$ K. Once the temperature reaches the sublimation temperature of H$_2$O, the whole ice mantle will
%evaporate.
%{\bf  Please state an example for this effect, perhaps by referring to Fig. 3.}

\section{Discussion}
\subsection{Effect of 2D structure}
So far, we have adopted a spherically-symmetric model of star-forming cores. In the spherical model,
fluid parcels fall to the central star over a short time scale.
But in reality, dense cores are rotating with $\omega\sim 10^{-14}$ s$^{-1}$, and the spherical symmetry
is broken in the central regions as the collapse proceeds. Because of  angular-momentum conservation,
the centrifugal force increases for infalling fluid parcels and eventually balances  the gravity
at the centrifugal radius, which is given by the expression
\begin{equation}
r_{\rm cent}=\frac{(r^2 \omega )_{\rm init}^2}{GM} \sim 100 \left(\frac{r}{0.07 {\rm pc}}\right)^4
\left(\frac{\omega}{1\times 10^{-14} {\rm s}^{-1}}\right)^2 \left(\frac{1M_{\odot}}{M}\right) {\rm AU}
\end{equation}
to form a circumstellar disk, where $r$ and $M$ are the size and mass of the initial dense core, respectively.
The result is  that the fluid parcels stay in  warm dense regions for
a longer time scale than occurs in the spherical model.

In order to investigate the chemistry in this prolonged warm phase, we have performed a 
pseudo-time-dependent model (i.e., fixed density and temperature). For the initial abundances, we adopt
the molecular abundances at 30 AU in the final step of the spherical collapse model ($t=9.3\times 10^4$ yr),
considering that the typical centrifugal radius is $\sim 100$ AU and that molecular abundances
at 2.5 -- 100 AU are almost constant in the collapse model at $9.3\times 10^4$ yr. All ices have already
evaporated to the gas phase in this initial abundances (Figure \ref{dist_abun}).

%{\bf RTG: I DON'T UNDERSTAND THE FINAL SENTENCE ABOVE.}

Figure \ref{disk} shows the temporal variation of molecular abundances at $T=40$ K ($a-b$)
and $T=150$ K ($c-d$). We choose these temperatures because grain-surface reactions would be
active at $T=40$ K, while gas-phase reactions would dominate at $T=150$ K.
The density is set to be $n_{\rm H}=1.487\times 10^8$ cm$^{-3}$, which
is the density at $r=30$ AU in the 1D model. These parameter choices are rather arbitrary, but
there should be regions with these densities and temperatures in the disks, because a real disk has
density and temperature gradients in both the vertical and radial directions. We simply assume that the
infalling fluid parcel reaches and stays in such regions in the disk. We also performed models with higher
($\times 10$) and lower ($\times 0.1$) densities, but the results are qualitatively the same; the
timescale of the chemical evolution described below depends only slightly on the density \citep{nomura09}.
At $T=40$ K, C$_2$H$_6$ increases and becomes the dominant carbon chain species. It is formed via
grain-surface hydrogenation of C$_2$H$_4$, which is produced by the gas-phase reaction of CH + CH$_4$.
It should be noted that C$_2$H$_4$ ($E_{\rm ads}= 3487$ K in our model; \citet{garrod06}) can marginally freeze onto grains,
while CH ($E_{\rm ads}= 925$ K) and CH$_4$ ($E_{\rm ads}=1300$ K) are easily desorbed
at 40 K; adsorption of C$_2$H$_4$ onto the grain surface works as a {\it sink} in the gas-phase chemical network.
Similarly, C$_7$H$_4$ ice ($E_{\rm ads}= 7487$ K) is the dominant carbon chain at $t=10^6$ yr at 150 K.
Since the sublimation temperature of C$_7$H$_4$ is close to 150 K, it depletes onto grains before additional
carbon-chain growth in the gas phase at high densities, while hydrogenation on grain surface is inefficient
at this high temperature.
%At $T=40$ K, carbon chains are adsorbed
%onto grains and hydrogenated 

% while at $T=150$ K, unsaturated carbon chains are formed efficiently
%in the gas phase {\bf although the radicals among them, such as CCH,  are depleted by reactions with H$_{2}$}.
The D/H ratios of the carbon chains remain high at $T=40$~K, since the
mother molecules (e.g. CH$_4$) are highly deuterated in the cold phase. Also, the exothermic exchange reactions
 CH$_3^+$ + HD and C$_2$H$_2^+$ + HD, which have higher exothermicities than H$_3^+$ + HD, 
contribute to keep the D/H ratios high.
At $T=150$ K, on the other hand, the D/H ratio of carbon-chain species
decreases gradually, because the endothermic exchange reactions (e.g. CH$_2$D$^+$ + H$_2$
and C$_2$HD$^+$ + H$_2$) can now proceed efficiently. It is interesting that species that undergo direct
exchange reactions with HD, such as H$_3^+$ and C$_2$H$_2^+$, work as a source of D/H enhancement in the network
at low temperatures, but easily lose their own D enhancements at high temperature. 

Among complex organic species, the HCOOCH$_3$ abundance increases from $1.6 \times 10^{-9}$ (in the gas phase) to
$\sim 10^{-7}$ (in the ice) via the grain-surface reaction of HCO + CH$_3$O in a few times $10^4$ yr
at $T=40$ K, while the abundances of other complex species do not change
significantly within this timescale.  At $T=150$ K, on the other hand, a significant amount of CH$_3$OCH$_3$
is formed in the gas phase from CH$_3$OH in $\sim 10^5$ yr. In spite of the formation at high
temperatures, the D/H ratios of CH$_3$OCH$_3$ is high, because the mother molecule,
CH$_3$OH, is highly deuterated in the cold phase.
At $t\gtrsim 10^5$ yr, complex organic species are destroyed via protonation and subsequent dissociative
recombinations.

Another important feature missing in our spherical model is outflows. It creates a cavity in the envelope,
through which the UV and X-ray from the protostar may escape. Photodissociation and photoionization play important
roles near the cavity wall \citep{visser11}, or the surface of the circumstellar disks, if the envelope mass is small.
The fraction of the mass going through such PDR-like regions is, however, rather
limited: $\lesssim$ several \% \citep{visser11}. Our 1D model would be a reasonable approximation for the rest of the
envelope regions, relocated from the outflow cavity.

\subsection{Comparison with observations}

%{\bf Yuri:   this section can be organized better.  I have tried my best to improve the discussion, but I am still confused on several aspects of the science.}

Table \ref{obs} lists estimated abundances with respect to H$_2$ from observations of oxygen-containing 
organic species in the protostellar core IRAS16293-2422. We distinguish the observations by single-dish telescopes
and interferometers. IRAS16293-2422 is a binary system; interferometric observations distinguish the two sources
IRAS16293A and IRAS16293B, while single-dish observations integrate the emission from the whole system.
As can be seen from the diverse results, it is not straightforward to 
estimate molecular abundances in protostellar cores,
because the cores have temperature and density gradients along the
line of sight, and because the high-temperature regions, in which these complex organic species sublime,
are mostly spatially unresolved even with the interferometer observations.
Abundance estimates therefore vary significantly depending on the assumed
temperature and density structure of the core.

It is noteworthy that interferometric observations tend to derive lower abundances of
large complex molecules than single dish observations corrected for beam dilution.
%It is unlikely that the lower abundances result from resolving out large structures,
%since the emissions are compact in single dish observations. Furthermore, the interferometric observations
%derive comparable or higher abundances of H$_2$CO and CH$_3$OH as compared with single dish observations.
\citet{oberg11} also noted such trends, and pointed out that cold extended component contributes significantly
to the single dish observation.
Thanks to the higher spatial resolution, interferometric
observations could derive more reliable abundances in the central hot ($\gtrsim 100$ K) regions.

Table \ref{obs} also lists the abundances in our model near the center ($r=30$ AU) of the
protostellar core at $t=9.3\times 10^4$ yr and in our pseudo-time-dependent model
with $T=40$ and 150 K at $t=10^5$ yr from the start of the ``disk". The core radius of 30 AU is
arbitrary, because the molecular abundances inside $\sim 100$ AU are almost constant due to rapid infall;
it simply represents the abundances at $T\gtrsim 100$ K.
As for the pseuo-time-dependent model at $T=40$ K, we list the ice abundances of complex organic species,
assuming that they eventually accrete to smaller radii to be sublimated to the gas phase.
As can be seen, the abundances of H$_2$CO and CH$_3$OH are higher in our core model than derived from
observation. In the pseudo-time-dependent models,
on the other hand, the H$_2$CO abundance is far too low at both temperatures
at the time used, while the methanol abundance for the 150 K model is in good agreement with observation.
Note though that the abundances of H$_{2}$CO and  CH$_{3}$OH  are time-dependent; these species decrease with time
at 150 K (Figure \ref{disk}).
We note that the spatial extent of the CH$_3$OH sublimation zone (i.e. $T\gtrsim 100$ K) in the protostellar core
is $\sim 100$ AU \citep{sjv02},
which is comparable to the typical centrifugal radius of dense cores. Therefore CH$_3$OH can be partially destroyed in
the prolonged warm temperature caused by rotational support. It is noteworthy that a recent ALMA observation
of methyl formate revealed a velocity gradient which is consistent with rotating gas around IRAS16293A, although it is not ruled out that the velocity
gradient could originate in two unresolved velocity components such as occur in a close binary \citep{pineda12}.
%Since the spatial extent of  gaseous CH$_3$OH  is  comparable with the centrifugal radius, the bulk
%amount of CH$_3$OH could be in the disk component rather than in the collapsing core.
Another explanation for the disagreement of CH$_3$OH and H$_2$CO abundances between our core model and
the observation might be that these abundances are underestimated by observations.
Observational constraints of molecule abundances in the warmer regions are uncertain because of beam dilution
effects, the presence of multiple systems, and the possible presence of cavities and outflows.
Indeed, interferometric observations which distinguish the two protostars in IRAS16293 system gives higher abundances
of CH$_3$OH than single dish observations.
%The low abundance of water  (2e-6 or less) observed in the hot corino of IRAS16293 by Crimier et al. (2010) may
%indicate that the source is more complex than 

Concerning the other molecules in the table, the model abundances
for HCOOCH$_3$, HCOOH, and CH$_3$CN in the core model are within the range of the estimated abundance from the
observations. CH$_3$OCH$_3$, on the other hand,  is significantly underestimated
in our core model. In the pseudo-time-dependent model at 150 K, however, the CH$_3$OCH$_3$ abundance increases with time
and agrees with the observation at $t\sim 10^5$ yr. It is formed by the recombination of CH$_3$OCH$_4^+$ in the
gas phase. We note that the branching ratio for producing CH$_3$OCH$_3$ in the recombination is only 5 \% in our model.
If we assume a higher branching ratio, CH$_3$OCH$_3$ becomes more abundant.
Abundances of other species in the pseudo-time dependent model at $10^5$ yr are also consistent with observations,
except for H$_2$CO.
This may indicate that the gases in the central hot region of IRAS16293 may have experienced prolonged warm
temperature chemistry compared with the spherical infall model due to partial rotational support.
Observations with higher angular resolution are highly desirable for more detailed comparison. 

Our core and pseudo-time-dependent models can also be applied to carbon-chain species.
\citet{nami10} observed carbon chains towards L1527, and showed that the intensity distribution of
C$_3$H$_2$ ($4_{3,2}-4_{2,3}$) is well reproduced, if its abundance jumps from $2.7\times 10^{-9}$
to $2.7\times 10^{-8}$ in the temperature range $T=23-29$ K. These peak abundances and temperature ranges are
in reasonable agreement with our core model at $t=9.3\times 10^4$ yr and $r\sim 10^3$ AU.
\citet{nami10} found that C$_2$H, C$_4$H, and C$_3$H$_2$ have a slight dip at the central region.
In our spherical  model, the central holes of the  C$_2$H and C$_4$H abundances are reproduced, but C$_3$H$_2$ increases
inwards, as discussed earlier.
In the pseudo-time-dependent model at 40 K, C$_3$H$_2$ is quickly depleted onto the grains, just like gaseous C$_2$H$_6$ in
Figure \ref{disk}. The central dip of C$_3$H$_2$ thus might be reproduced, if the central region is dominated by the
cold disk material rather than a warm infalling envelope. It is possible that the disk midplane is much cooler than
the envelope because of the higher optical depth.

Let us now move on to deuterated species.  Table \ref{obs_DH} lists the estimated molecular D/H ratios in various 
protostellar cores and our model
D/H ratios. As in Table \ref{obs}, we list model D/H ratios at $t=9.3\times 10^4$ yr of the
protostellar core at $r=30$ AU, except for C$_4$H, and in our pseudo-time-dependent models.
For the C$_4$D/C$_4$H ratio in the core model, we list the ratio at $r=250$ AU, where C$_4$H abundance reaches its maximum.
Our D/H ratios for the carbon chain species C$_{4}$H and C$_{3}$H$_{2}$, and for NH$_3$, are in reasonable
agreement with the observations of L1527.
On the other hand, the observed D/H ratios of NH$_3$ in NGC1333 and H$_2$CO in several protostellar cores are much higher
than in our models (either core or disk).
It should be noted, however, that in our core model these ratios are higher in the central warm region
at early stages (e.g. $t=4.3\times 10^2$ yr) and in the CO depletion zone than listed in Table 3 (Figure \ref{dist_multiD}).
The model D/H ratios vary significantly depending on
the evolutionary stage of the core, and the size of the warm regions, while the observational D/H ratios thus
depend on beam size
(whether the beam traces mainly the cold component or the central warm regions).
Such dependence also explains the variety of observed D/H ratios towards protostars.

The CH$_2$DOH/CH$_3$OH ratio reaches a high of 0.65 in the observations, while it is only a few per cent in
our models. The comparison is even worse for CHD$_2$OH/CH$_3$OH.
Unlike NH$_3$ and H$_2$CO, CH$_3$OH is not formed in the gas phase, and its gaseous abundance is very low
in the CO depletion zone. Hence it is unlikely that the high CH$_2$DOH/CH$_3$OH ratio originates in the CO depletion zone.
In our model, CH$_2$DOH is formed by the addition of D during the hydrogenation of CO on grain surfaces;
we have not included substitution reactions of H$_2$CO + D, or the abstraction reaction CH$_3$OH + D, which are
observed in laboratory experiments \citep{nagaoka05,nagaoka07,hidaka09}. \citet{taquet12} recently performed
gas-grain chemical models at constant density and temperatures, and showed that
these reactions are crucial to reproduce the observed D/H ratios of H$_2$CO and CH$_3$OH.
In the following subsection (\S4.3) , we add the substitution and abstraction reactions on grain surfaces to our network
to see how strongly the CH$_2$DOH/CH$_3$OH abundance ratio is enhanced.

The D/H ratio of water ice in the envelope at $t=9.3\times 10^4$ yr is $\sim 1$ \%, which is marginally consistent
with the upper limits obtained towards low-mass protostars \citep{parise03}.
Recently gaseous HDO has also been intensively searched for towards protostellar cores using ground-based single-dish telescopes
and the Herschel Space Observatory. Table \ref{obs_HDO} lists the estimated HDO abundance and HDO/H$_2$O ratio in the gas phase
in the hot corino region ($T\ge 100$ K) and outer envelope ($T\le 100$ K) towards NGC 1333-IRAS2A and IRAS16293
\citep{liu11, cousten12}.
Towards IRAS16293, an absorbing layer is added in front of the envelope to account for the observed absorption lines.
Our model HDO/H$_2$O ratio in the hot region is consistent with the observations, while our HDO abundance is
higher by an order of magnitude (see panel ($a$) of $t=9.3\times 10^4$ yr in Figure 3).
We note that the HDO (and H$_2$O) abundance could be underestimated in observations,
just like CH$_3$OH and H$_2$CO, because of the complex (and poorly-constrained) structures of the source such as the
presence of multiple systems (IRAS16293), and the possible presence of cavities, outflows and disks.
In addition to the targets listed in Table \ref{obs_HDO}, \citet{jorgensen10} searched for gaseous HDO towards NGC1333-IRAS4B
using the Sub-Millimeter Array (SMA). The line was not detected, which constrains the HDO/H$_2$O ratio $\le 6\times 10^{-4}$ in the central hot
region ($\sim 50$ AU). This upper limit is much lower than predicted in our model. Since gaseous HDO in the central region
originate in ice formed in the cold era, the variation of the HDO/H$_2$O ratio among
protostars could originates in the temperature variation in the very early stages of molecular clouds or translucent clouds
\citep{cazaux11}.

In the outer envelope, our HDO abundance and HDO/H$_2$O ratio at $\gtrsim 10^3$ AU at $t=9.3\times 10^4$ yr are
in reasonable agreement with NGC1333-IRAS2A and absorbing layer of IRAS16293 (Figure 3).
But the observed values in the envelope
of IRAS16293 is lower than in our model at $t=9.3\times 10^4$ yr. Since the  molecular column densities in the envelope
are lower in earlier stages, a model at slightly earlier stage than $t=9.3\times 10^4$ yr might be more
consistent with the envelope HDO abundance in IRAS16293.
It should also be noted that in our model at
$t=9.3\times 10^4$ yr, HDO/H$_2$O ratio is close to unity around 500 AU, which is a part of the outer envelope
(Figure \ref{dist_DH}). This high HDO/H$_2$O ratio is caused by the exothermic exchange reaction of OH + D;
the rate coefficient is set to $k=1.3\times 10^{-10}$ cm$^3$s$^{-1}$ \citep{millar89}, which is consistent
with a detailed theoretical calculation by \citet{atahan05} ($1.5 \times 10^{-10}$ cm$^3$s$^{-1}$ at 50 K).
Such local HDO/H$_2$O enhancement has not been observed thus far.
Although the effect of the OH + D reaction seems to be transient and local in our model, a further search for such
enhancements could constrain O-chemistry in the protostellar envelope.

%We note, however, that this local high HDO/H$_2$O ratio (and high HCOOD/HCOOH ratio) disappears, and the ratio in the
%outer envelope becomes more consistent with the observed value, if we include
%the D-H exchange reaction of CH$_3$OH on the grain surface (see \S 5.2), which consumes a significant amount of D atoms.

\begin{table}
\caption{ Observed and Model Gas-Phase Molecular Abundances in IRAS 16293-2422}
\begin{center}
\begin{tabular}{l c c c c c}
\hline
Species & \multicolumn{2}{c}{IRAS 16293-2422} & \multicolumn{3}{c}{model\tablenotemark{a}}\\
      & single dish  & interferometer  & core & 40 K & 150 K\\
\hline
%H$_2$CO & 6.0(-8)\tablenotemark{b}, 1.1(-7)\tablenotemark{c} \\
%CH$_3$OH & 3.0(-7)\tablenotemark{d}, 9.4(-8)\tablenotemark{c} \\
H$_2$CO & 1.0(-7)\tablenotemark{b} & 1.1(-7)\tablenotemark{c} & 1.1(-5) & 5.4(-10)\tablenotemark{d}& 1.1(-11)\\
CH$_3$OH & 1.0(-7)\tablenotemark{e} & 9.4(-8)\tablenotemark{c}, 6.8(-7)\tablenotemark{f}, 3.1(-7)\tablenotemark{g} 
& 3.6(-6) & 2.9(-6)\tablenotemark{d} & 5.2(-7)\\
HCOOCH$_3$ &  4.0(-7)\tablenotemark{h} &  7.5 (-9) \tablenotemark{i},4.3(-9)\tablenotemark{f}, 2.6(-9)\tablenotemark{g}& 3.1(-9) & 1.7(-7)\tablenotemark{d}& 8.8(-10)\\
           & & 4.9 (-10)\tablenotemark{j},$>$ 1.2(-8)\tablenotemark{k} & & \\ 
HCOOH   & 6.2(-8)\tablenotemark{h} & 2.5(-9)\tablenotemark{k} & 1.1(-8) & 3.3(-8)\tablenotemark{d} & 1.2(-8)\\
CH$_3$OCH$_3$ &  2.4(-7)\tablenotemark{h} & 7.6(-8)\tablenotemark{c} & 1.2(-10) & 3.9(-10)\tablenotemark{d}& 3.3(-8)\\
CH$_3$CN & 7.5(-9)\tablenotemark{l}, 1.0(-8)\tablenotemark{h} &4.8 (-10)\tablenotemark{i},  2.3(-11)\tablenotemark{j}  & 5.6(-9) & 1.3(-8)\tablenotemark{d}&4.3(-9)\\
\hline
\end{tabular}
\end{center}
\tablenotetext{a}{Molecular abundances at $r=30$ AU and $t=9.3\times 10^4$ yr in the
protostellar core model, and in the pseudo-time-dependent model
of $T=40$ K and 150 K at $t=1\times 10^5$ from the start of the "disk".}
\tablenotetext{b}{\citet{maret04}}
\tablenotetext{c}{\citet{cha05}}
\tablenotetext{d}{ice abundance}
\tablenotetext{e}{\citet{maret05}}
\tablenotetext{f}{\citet{kua04}, IRAS16293A}
\tablenotetext{g}{\citet{kua04}, IRAS16293B}
\tablenotetext{h}{\citet{caz03}}
\tablenotetext{i}{IRAS16293 A \citep{bottinelli04}}
\tablenotetext{j}{IRAS16293 B \citep{bottinelli04}}
\tablenotetext{k}{\citet{rem06}}
\tablenotetext{l}{\citet{sjv02}}

\label{obs}
\end{table}%

\begin{table}
\caption{Molecular D/H ratios in protostellar cores}
\begin{center}
\begin{small}
\begin{tabular}{l c c c c l l }
\hline
Species & D/H ratio & source & \multicolumn{3}{c}{model\tablenotemark{a}}& reference\\
        &           &        & core & 40 K & 150 K &\\
\hline
%H$_2$CO & 6.0(-8)\tablenotemark{b}, 1.1(-7)\tablenotemark{c} \\
%CH$_3$OH & 3.0(-7)\tablenotemark{d}, 9.4(-8)\tablenotemark{c} \\
NH$_2$D/NH$_3$ & 0.28 & NGC1333 & 0.015 & 0.014\tablenotemark{b} & 0.012 &\citet{roueff05}\\
               & 0.04 & L1527 &  & & & \citet{nami09}\\
ND$_3$/NH$_3$ & $2.1\times 10^{-4}$ & NGC1333 & $1.6\times 10^{-6}$ & $1.5\times 10^{-6}$\tablenotemark{b} & $5.1\times 10^{-7}$ & \citet{roueff05}\\
HDCO/H$_2$CO & 0.094-1.7 & 7 protostars & 0.026 & 0.013\tablenotemark{b}& -\tablenotemark{c} &\citet{parise06}\\
D$_2$CO/H$_2$CO & 0.046-0.44 & 7 protostars & $3.3\times 10^{-5}$ & $2.9\times 10^{-5}$\tablenotemark{b}  
 &  -\tablenotemark{c} &\citet{parise06} \\
CH$_2$DOH/CH$_3$OH & 0.37-0.65 & 4 protostars & 0.024 & 0.026\tablenotemark{b} & 0.023 &\citet{parise06}\\
                   & $\le 0.030$ & L1527 & & & &\citet{nami09}\\
CHD$_2$OH/CH$_3$OH & 0.074-0.25 & 4 protostars &  $1.7\times 10^{-4}$ & $2.0\times 10^{-4}$\tablenotemark{b}
& $1.5\times 10^{-4}$  & \citet{parise06}\\
C$_4$D/C$_4$H & 0.018 & L1527 & 0.033\tablenotemark{d} & -\tablenotemark{c}& -\tablenotemark{c} & \citet{nami09}\\
C$_3$HD/C$_3$H$_2$ & 0.071 & L1527 & 0.033 & 0.016\tablenotemark{b} & 0.0083 &\citet{nami09}\\
\hline
\end{tabular}
\end{small}
\end{center}
\tablenotetext{a}{Molecular abundances at $r=30$ AU and $t=9.3\times 10^4$ yr in the
protostellar core model, and in the pseudo-time-dependent model
of $T=40$ K and 150 K at $t=1\times 10^5$ from the start of the "disk".}
\tablenotetext{b}{ice abundance}
\tablenotetext{c}{D/H ratio is not listed because the abundance of the molecule is too small compared with observations.}
\tablenotetext{d}{D/H ratio at $r=250$ AU, where C$_4$H abundance reaches its maximum.}
\label{obs_DH}
\end{table}%

\begin{table}
\caption{ HDO abundance and HDO/H2O ratio in the gas phase in protostellar cores}
\begin{center}
\begin{small}
\begin{tabular}{l c c c l}
\hline
    & \multicolumn{3}{c}{regions}& reference\\
    & hot ($T\ge 100$ K) & envelope ($T\le 100$ K) & absorbing &\\
\hline
NGC1333-IRAS2A & & & & \citet{liu11} \\
HDO abundance & $6.6\times 10^{-8}-1.0\times 10^{-7}$ & $9\times 10^{-11}-1.8\times 10^{-9}$& - & \\
HDO/H$_2$O & $\ge 1$ \% & $0.9- 18$ \% & - & \\
IRAS16293-2422 & & & & \citet{cousten12}\\
HDO abundance & $1.7\times 10^{-7}$ & $8\times 10^{-11}$ & \\
HDO/H$_2$O & $1.4-5.8$ \% & $0.2-2.2$ \% & 4.8 \% \\
\hline
\end{tabular}
\end{small}
\end{center}
\label{obs_HDO}
\end{table}%

\subsection{Deuteration of CH$_3$OH on grain surfaces}
The grain-surface reaction network for H$_2$CO and CH$_3$OH derived from laboratory experiments is summarized
in \citet{watanabe08} and
\citet{hidaka09}. Most of the reactions are already included in our original gas-grain model, except for the
substitution H$_2$CO + D  $\rightarrow$ HDCO + H, abstraction CH$_3$OH + D $\rightarrow$ CH$_2$OH + HD, and their
deuterated analogues (Figure \ref{Eact}). Here we add these reactions to our model.

Laboratory experiments indicate that the abstraction reaction CH$_3$OH + D produces CH$_2$OH but not CH$_3$O, so
the subsequent hydrogenation reaction can form CH$_2$DOH but not CH$_3$OD, which is in agreement with the observed
low abundance of CH$_n$D$_{3-n}$OD compared with CH$_n$D$_{3-n}$OH \citep{parise04, nagaoka05}.
In our original deuterated network model, however, we did not follow the position of the deuterium atoms in the reactions,
and we did not fully distinguish the various possible isomers of radicals, such as CH$_2$DO, CHDOH and CH$_2$OD, which
are produced in the sequence of deuterated methanol formation. Our model thus underestimates (dilutes) the
isomer ratio CH$_n$D$_{3-n}$OH/CH$_n$D$_{2-n}$OD, although we distinguish isomers in the abstraction reactions if they are
in our original species list. In spite of the imperfect distinction of the isomers,
CH$_2$DOH is more abundant than CH$_3$OD in our model.

The activation barriers of the substitution and abstraction reactions have recently been investigated using theoretical
calculations \citep[e.g.][]{goumans11}; however, the rate is determined not only by activation barrier height. Since the reactions proceed via tunneling
at low temperatures, the tunneling mass and the shape (width) of the potential barrier are also important.
Thus, the accurate evaluation of
reaction rates from theoretical calculations is challenging. Rather than incorporate activation barriers from theoretical calculations, here we utilize the rates derived from the laboratory experiments by \citet{watanabe08} and
\citet{hidaka09}, who estimated relative rate coefficients of various reactions to the rate of H + CO.
In our model, rate coefficients for grain-surface reactions are proportional to a frequency factor times
$\exp [-{2w/\hbar}\sqrt{2m E_{\rm act}}]$, where $w$ is a width of the barrier, $m$ is the effective mass,
and $E_{\rm act}$ is the height of the barrier, if the reaction proceeds via tunneling.
The width of the barrier $w$ is fixed to 1 \AA.
The effective mass is equal to the reduced mass of the reactants in a two-body addition reaction, but depends
on the details of reaction for more complex reactions such as abstraction \citep{hidaka09}.
Since our model includes various
grain-surface reactions of which the details (such as potential energy surface) are not yet well understood,
we use the reduced mass of the reactants as the effective mass for all grain-surface reactions. 
The activation barrier of the reaction H + CO is
set to 2500 K in our original model. Figure \ref{Eact} shows reactions in our model and $E_{\rm act}$ which we assume
for each reaction in units of K;
$E_{\rm act}$ is determined so that the rate coefficients relative to H + CO become roughly 
consistent with \citet{watanabe08} and \citet{hidaka09}. Reactions depicted by the solid arrows are included in our
original model, while the dashed arrows are newly added reactions. In the original model, the activation barrier
for D + CO is set to be the same as H + CO (2500 K); due to the larger effective mass, the rate coefficient of D + CO
is lower than that of H + CO by several orders of magnitude. In the laboratory experiment, on the other hand, 
the relative rate coefficient of D + CO is about 10 \% to H + CO. Hence we changed the barrier for D + CO to
1400 K in the model. Similarly the barrier for reaction of D atom with H$_2$CO (and its deuterated isotopomers) are
lowered.
It should be noted that these values of $E_{\rm act}$ are not necessarily equal to the activation barrier
obtained from the quantum chemical calculations, since the rate of the tunneling sensitively depends on the shape of
the potential and effective mass in reality, while we assume a simple rectangular barrier and adopt the reduced mass
as the effective mass.
We also note that the frequency factor is often replaced by accretion or desorption rates of
H and D atoms in our calculation, because we adopt the modified rate.

\citet{taquet12} calculated the D/H ratios of H$_2$CO and CH$_3$OH in pseudo-time-dependent models of
$n_{\rm H}=10^4-5\times 10^6$
cm$^{-3}$ cm$^{-3}$, including the substitution and abstraction reactions. Their grain-surface chemistry model
is more sophisticated than ours; they consider multiple layers and cracks in the grain mantles.
The rate coefficients are determined relative to the H + CO rate, referring to \citet{watanabe10} and \citet{hidaka09}.
They showed that the D/H ratios reach higher values at higher densities; in the model with $n_{\rm H}=5\times 10^6$
cm$^{-3}$ and $T=10$ K, the CH$_2$DOH/CH$_3$OH ratio exceeds unity at $t\gtrsim 10^4$ yr.

In order to compare our chemical network model with the work of \citet{taquet12}, we calculated a
pseudo-time-dependent model with $n_{\rm H}=5\times 10^6$ cm$^{-3}$ and $T=10$ K;
Figure \ref{cloud_CH3OH} shows the temporal variation of H$_2$CO, CH$_3$OH and their deuterated isotopes in our model.
The initial abundances
are set by solving the network under molecular-cloud conditions ($n_{\rm H}= 2\times 10^4$ cm$^{-3}$, $T=10$ K)
for $8\times 10^4$ yr in panel (a) and for $2\times 10^5$ yr in panel (b). The molecular D/H ratios become very high
$0.1-1.0$ at several $10^4 -10^5$ yr. The rise in the D/H ratio is slower than in the model of
\citet{taquet12}, in which the HDCO/H$_2$CO ratio exceeds 0.1 at $\sim 10^3$ yr and the CH$_2$DOH/CH$_3$OH
ratio exceeds unity at $\sim 10^4$ yr. The slow rise of the D/H ratio in our model would be mainly due to the
initial abundances. \citet{taquet12} calculate the
atomic D/H ratio in steady state as a function of density, and adopt it as an initial abundance. The steady-state
atomic D/H ratio is greater under higher-density conditions. Our initial abundances, on the other hand, are determined by the molecular-cloud calculation of $n_{\rm H}=2\times 10^4$ cm$^{-3}$.
In our pseudo-time-dependent model of $n_{\rm H}=5\times 10^6$ cm$^{-3}$, the atomic D/H ratio
increases in $10^4$ yr to reach the steady state.
Initial abundances of CO, CH$_3$OH and H$_2$CO would also be relevant; the temporal variation of the molecular
D/H ratio differs
in panel (a) and (b). Although deuterated H$_2$CO and CH$_3$OH increase rapidly both in (a) and (b) at $t\gtrsim 10^4$ yr,
when the atomic D/H ratio reaches steady state, the molecular D/H ratios do not reach unity in panel (b), due to the
high abundance of normal isotopes in the initial conditions. The model of \citet{taquet12}, on the other hand, assumes that all carbon initially resides in CO.

Having tested our chemical network against \citet{taquet12}, we now calculate 
molecular abundances and D/H ratios 
including the substitution and abstraction reactions at $t=-5.6\times 10^2$ yr, 
$4.3\times 10^2$ yr and $9.3\times 10^4$ yr in our star-forming cores.
The result is almost the same as in Figure 3; the D/H
ratio increases at most by a factor of a few. There are three possible reasons for this low D/H ratio in spite
of the newly-added reactions:
\begin{itemize}
\item[(i)]{Our model starts from a dense molecular cloud core. A significant amount of CH$_3$OH and H$_2$CO is
already formed during the pre-collapse phase with $n_{\rm H}\lesssim 5\times 10^4$ cm$^{-3}$, during which
the deuterium enrichment by the new reactions are not very efficient \citep{taquet12}.}
\item[(ii)]{The period spent at high density ($\gtrsim 10^6$ cm$^{-3}$), in which the deuterium enrichment
is efficient, is limited by the dynamical model. The temporal variation
of density in the infalling fluid parcels that reach $R=2.5$ AU at $9.3\times 10^4$ yr is shown in Figure 1.
The parcel falls onto the protostar in several $10^3$ yrs after it enters the region where
$n_{\rm H}\ge 10^6$ cm$^{-3}$.}
\item[(iii)]{The duration of the high-density phase is longer in fluid parcels that reach the central region
in earlier stages (e.g. $t=-5.6\times 10^2$ yr and $4.3\times 10^2$ yr); but the dust temperature for such parcels
is mostly very low; it could reach $\sim 5$ K \citep[e.g.][]{keto10}.
Since we assume thermal hopping, even H and D atoms cannot efficiently migrate on grains.}
\end{itemize}
Our model indicates that the D/H ratios of CH$_3$OH and H$_2$CO in the star-forming core
depend significantly on the dynamical model,
the heat balance (e.g. cosmic-ray flux) inside the prestellar cores, and the efficiency of grain-surface
migration of H and D atoms at very low temperature. For example, we found that the gaseous CH$_2$DOH/CH$_3$OH ratio
reaches 17 \% inside the sublimation radius ($r\le 7.5$ AU) at $t=4.3\times 10^2$ yr,
if we allow H and D atoms to migrate via tunneling.
Recent laboratory work excludes the migration of H and D atoms via tunneling, but finds fast thermal hopping at 8 K
\citep{watanabe10}.
Considering the strong dependence of the hopping rate on temperature, thermal hopping of H and D atoms could
still be inefficient at 5 K. In such a case, the Eley-Rideal mechanism, which is not included in our model, could be important.
We postpone further investigation of this issue to future work.

\section{Summary}
We have investigated molecular abundances and D/H ratios in a star-forming core by adopting the
1D radiation hydrodynamics model of \citet{masunaga00}.

Spatial distributions of assorted molecules in the gas phase and ice mantle
are reported for a core at a time $5.6\times 10^2$ yr prior to protostar formation (at which the so-called
first core is formed), and at $4.3\times 10^2$ yr and $9.3\times 10^4$ yr after the protostar is
born. Significant depletion onto grains occurs in the early stages, at radii 10 AU $\lesssim r \lesssim 10^3$ AU and 100 AU $\lesssim r \lesssim 10^3$ AU, for times $t=-5.6\times 10^{-2}$ yr and $4.3\times 10^2$ yr, respectively.
As the core gets warmer, complex species, as well as simple species such as NH$_3$, 
sublime from the grain surfaces near the core center. The abundances of these complex
species -- both in the gas phase and ice mantle -- increase with time, while the variation is rather
small for CH$_3$CN, which can be formed readily at low temperatures. Methanol, which can also be formed
at low temperatures, decreases gradually as the core evolves.
We find reasonable agreement with observation for 
our model abundances at $t=9.3\times 10^4$ yr and $T\gtrsim 100$ K, except for CH$_3$OCH$_3$, which
is significantly under-abundant in our 1D core model.

Unsaturated carbon chains are formed
from sublimed CH$_4$ at $T\sim 25$ K ($r\sim 1000$ AU) at $t=9.3 \times 10^4$ yr. 
We show that the chemistry that forms these chains is indeed the WCCC mechanism developed by
\citet{sakai08} to explain the chemistry in L1527 and IRAS15398-3359.
Our model also reproduces the central dip for C$_2$H and C$_4$H observed by \citet{nami10}.
In the earlier stages,
however, the gas-phase formation of unsaturated carbon chains is much less efficient due to the higher density
at the CH$_4$ sublimation zone. In order for WCCC to be efficient, the gas density at the CH$_4$
sublimation zone should be relatively low, so that C$^+$, which is another ingredient for carbon chain formation,
 becomes abundant.
%In this early phase, simple sublimation of carbon chains (such as C$_3$H$_2$)
%overwhelms the gas-phase formation of carbon chains in the central regions.

In the cold phase ($T\sim 10$ K), molecular D/H ratios are enhanced by  exothermic exchange reactions
such as H$_2$D$^+$ + HD. We find very high molecular D/H ratios (close to unity)
for gas-phase species in the CO depletion zone, because CO is the main reaction partner of H$_2$D$^+$.
In the central regions of the protostellar core, however, molecular D/H ratios are mostly $\sim 10^{-2}$; they
originate in the D/H ratios formed on ice mantles, which are accumulated  from the cloud core phase
(without significant CO depletion) all the way through to the protostellar phase. At $t=9.3\times 10^4$
yr, a significant amount of complex organic species and carbon chains are formed via grain-surface and gas-phase
reactions at warm temperatures ($T\gtrsim 30$ K for organic species and $T\gtrsim 25$ K for carbon chains).
Their D/H ratios are $\sim 10^{-2}$, inheriting the D/H ratio of their precursor molecules, and vary slightly among species.
Exothermic exchange reactions are still active at temperatures of a few tens of Kelvin, although the rates
of the backward reactions increase with temperature.
HCOOH has a higher D/H ratio than other complex organic species;
it is formed by H$_2$CO  + OH, so that HCOOD can be formed from H$_{2}$CO + OD
where OH is deuterated by the exchange reaction of OH + D. This exchange reaction also significantly enhances
the HDO/H$_2$O ratio at $T\sim 50$K. 
%The small observed HDO/H$_2$O ratio (a few \%) towards IRAS16293 is better
%reproduced in a model without this exchange reaction.

While the observed molecular D/H ratios in protostellar cores could be mostly explained by the combination
of D/H ratios in the hot-corino region (i.e. $T>100$ K) and CO depletion zone,
the D/H ratio of CH$_3$OH are significantly underestimated
in our model. Laboratory experiments \citep{nagaoka05, watanabe08, hidaka09} found substitution and abstraction reactions
on grain surfaces to deuterate H$_2$CO and CH$_3$OH; we have added these reactions to our model. Although the
D/H ratios of CH$_3$OH and H$_2$CO can exceed unity at $t\gtrsim 10^5$ yr in a pseudo-time-dependent model
with a high density ($n_{\rm H}=5\times 10^6$ cm$^{-3}$) and low temperature ($T=10$ K),
the ratios increased only by a factor of a few in our protostellar core model. A significant amount of CH$_3$OH
ice is formed in the prestellar stage, in which the density is still too low for the newly-added reactions to
be efficient. While the high density phase lasts longer in infalling fluid parcels in earlier evolutionary stage of
a core, the temperature could be as low as 5 K there. Eley-Rideal mechanism, rather than Langmuir-Hinshelwood
mechanism assumed in our model, could therefore be important. Further studies are needed on this issue.

Finally, in order to see how the molecular abundances evolve after fluid parcels land on the disk,
we performed pseudo-time-dependent models at $T=40$ K and 150 K starting from the molecular abundances set by
the 1D collapse model. At $T=40$ K, a significant amount of HCOOCH$_3$ ice is formed by the
grain-surface reactions. C$_2$H$_6$ ice becomes abundant among carbon chains;
it is formed via grain-surface hydrogenation of C$_2$H$_4$, which is produced by a gas-phase reaction.
Since C$_2$H$_4$ can marginally freeze onto grains at 40 K, its adsorption onto grains works as a sink
in the gas-phase chemical reaction network.
At $T=150$ K, C$_7$H$_4$ ice becomes the dominant carbon chain, and its D/H ratio
decreases with time at $t\gtrsim 10^5$ yr.
Meanwhile, CH$_3$OCH$_3$ is efficiently formed in the high-temperature gas from the sublimed CH$_3$OH. It is noteworthy
that  CH$_3$OCH$_3$ has a high D/H ratio ($\sim 0.040$), in spite of its formation at  high temperature,
because its mother molecule, CH$_3$OH, is highly deuterated during the cold phase.
It is also interesting that CH$_3$OCH$_3$ is under-abundant in our core model compared with observation,
while its higher abundance in the pseudo-time dependent model is in much better agreement. The pseudo-time-dependent
models also give higher abundances for some other complex organic species such as HCOOCH$_3$ compared with those
in the 1D core model.
Since the size of the hot corino region ($T\ge 100$ K) is close to a typical centrifugal radius ($\sim 100$ AU) of
cloud cores, at least a fraction of the large organic species might be formed in such prolonged
(i.e. longer than free-fall time scale) warm chemistry due to partial rotational support.
Such predictions should be tested via observations at high spatial resolution.

\acknowledgments
We would like to thank the anonymous referee for useful comments to improve our manuscript.
We are grateful to Prof. Naoki Watanabe for helpful discussions on grain-surface reactions.
This work was supported by  Grant-in-Aid for Scientific Research (A) 21244021, (C) 23540266, and
Grant-in-Aid for Scientific Research on Innovative Areas 23103004. VW and FH thank the French
CNRS/INSU program PCMI for its financial support.  EH thanks the National Science Foundation (US) for his program in astrochemistry and NASA for the study of pre-planetary matter.
Some kinetic data used here has been
downloaded from the online database KIDA (KInetic Database for Astrochemistry,
http://kida.obs.u-bordeaux1.fr).  


%% in parentheses, after the keyword, but they will not be verified.



\clearpage

%% the file, uncomment the \plotone, \plottwo, and
%% If you need a layout that cannot be achieved with \plotone or
%% \plottwo, you can invoke the graphicx package directly with the
%% \includegraphics command or use \plotfiddle. For more information,
%% please see the tutorial on "Using Electronic Art with AASTeX" in the
%% documentation section at the AASTeX Web site,
%% http://www.journals.uchicago.edu/AAS/AASTeX.
%%
%% The examples below also include sample markup for submission of
%% supplemental electronic materials. As always, be sure to check
%% the instructions to authors for the journal you are submitting to
%% for specific submissions guidelines as they vary from

%% This example uses \plotone to include an EPS file scaled to
%% 80% of its natural size with \epsscale. Its caption
%% has been written to indicate that additional figure parts will be
%% available in the electronic journal.

\clearpage

%% Here we use \plottwo to present two versions of the same figure,
%% one in black and white for print the other in RGB color
%% for online presentation. Note that the caption indicates
%% that a color version of the figure will be available online.
%%

\begin{figure}
\plotone{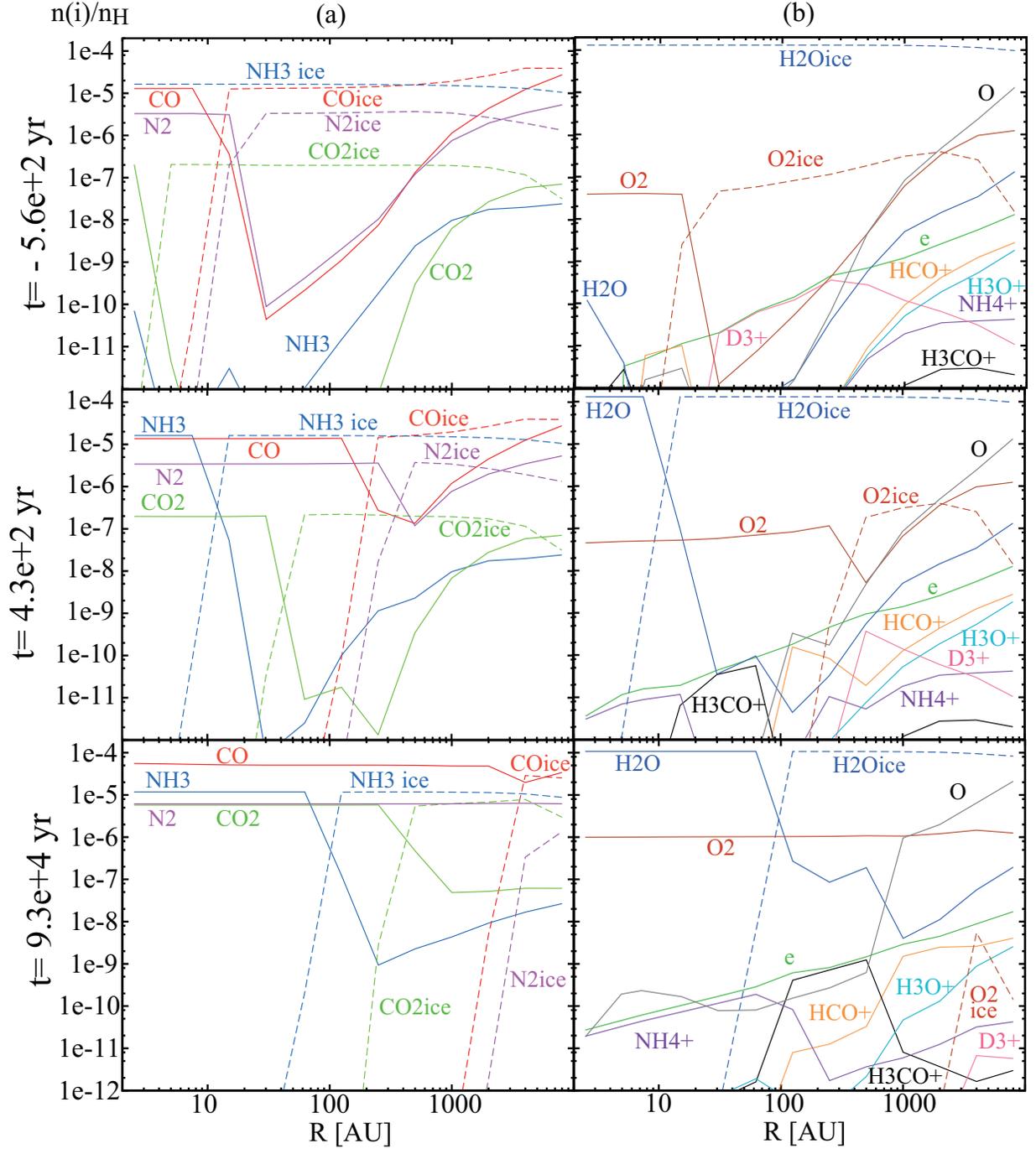}
\caption{Radial distribution of simple molecules ($a-b$), complex organic species ($c$),
and carbon chain species ($d$) at $t=-5.6\times 10^2$ yr, $4.3\times 10^2$ yr, and
$9.3\times 10^4$ yr. Solid lines and dashed lines depict gas-phase species and ice mantle species, respectively,
while the chemical species are distinguished by line colors.
Note that the horizontal axis starts from 2 AU rather than 1 AU.
\label{dist_abun}}
\end{figure}

\setcounter{figure}{1}
\begin{figure}
\plotone{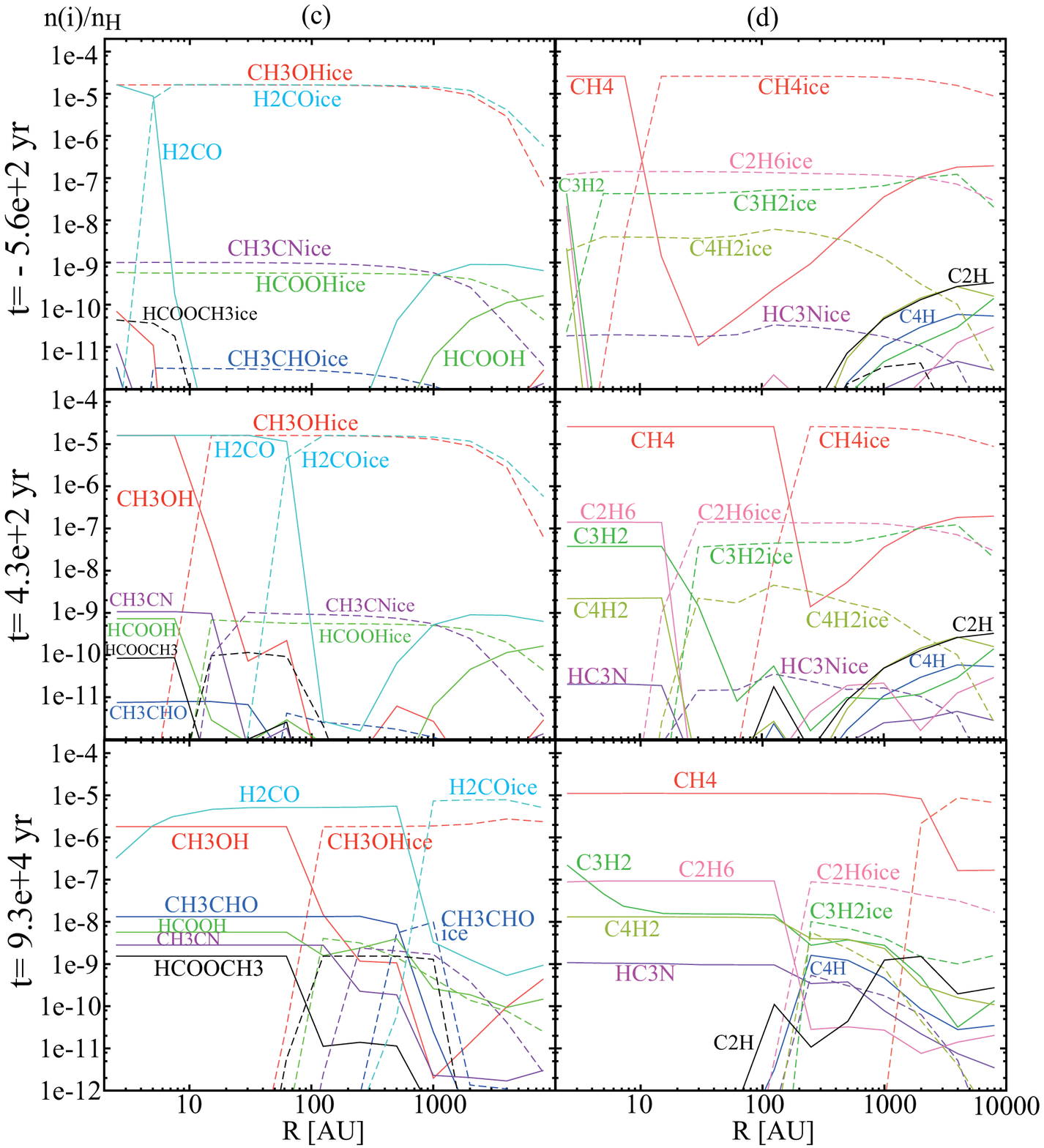}
\caption{(cont.)}
\end{figure}

\begin{figure}
\epsscale{0.8}
\plotone{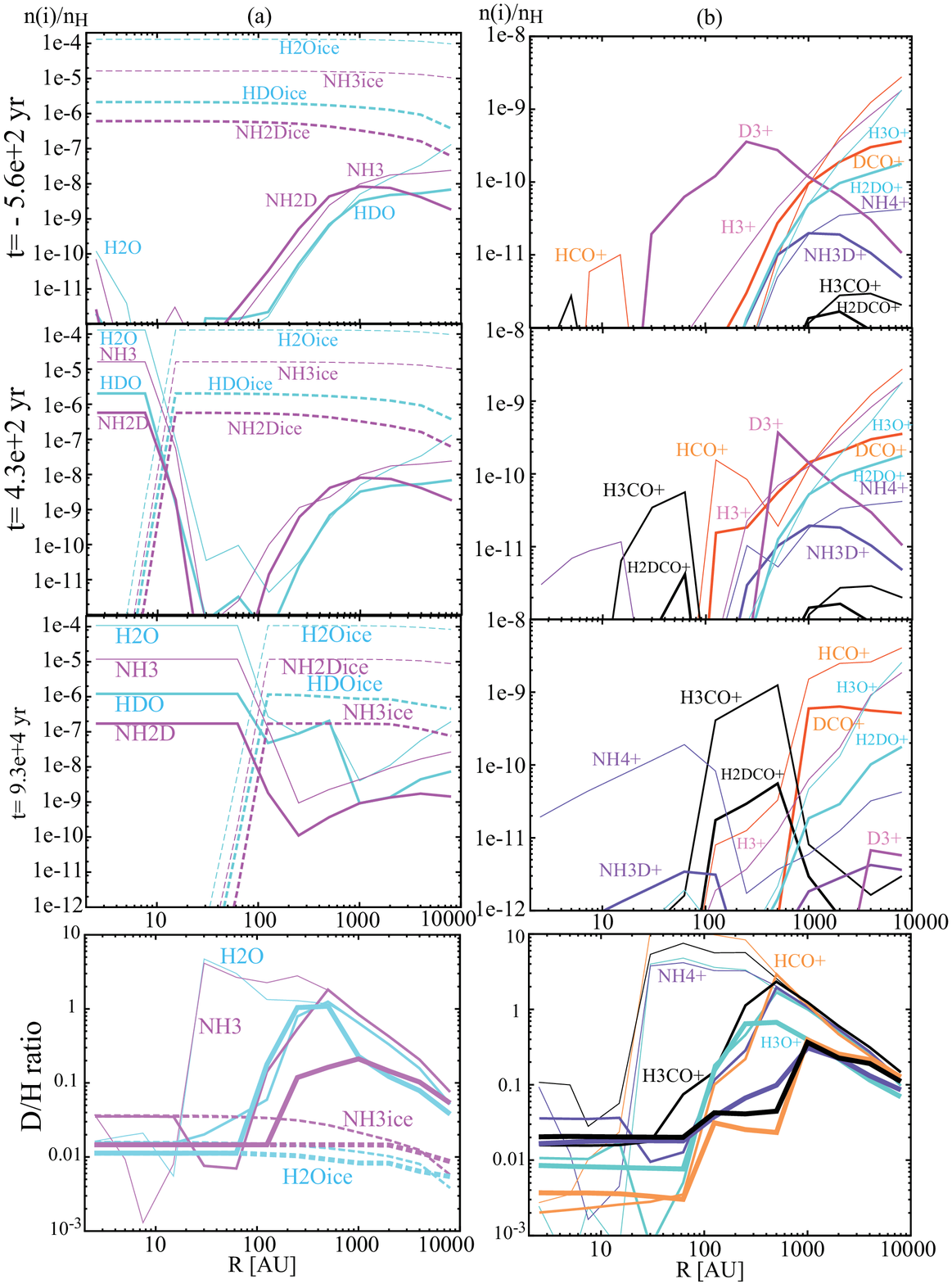}
\caption{Radial distribution of mono-deuterated species and their normal isotope
counterparts at $t=-5.6\times 10^2$ yr, $4.3\times 10^2$ yr, and
$9.3\times 10^4$ yr. The bottom panels show the radial distribution of  the D/H ratio for assorted
molecules. The thin, thick, and bold lines depict the D/H ratios at $t=-5.6\times 10^2$ yr,
$4.3\times 10^2$ yr, and $9.3\times 10^4$ yr, respectively.
\label{dist_DH}}
\end{figure}

\setcounter{figure}{2}
\begin{figure}
\plotone{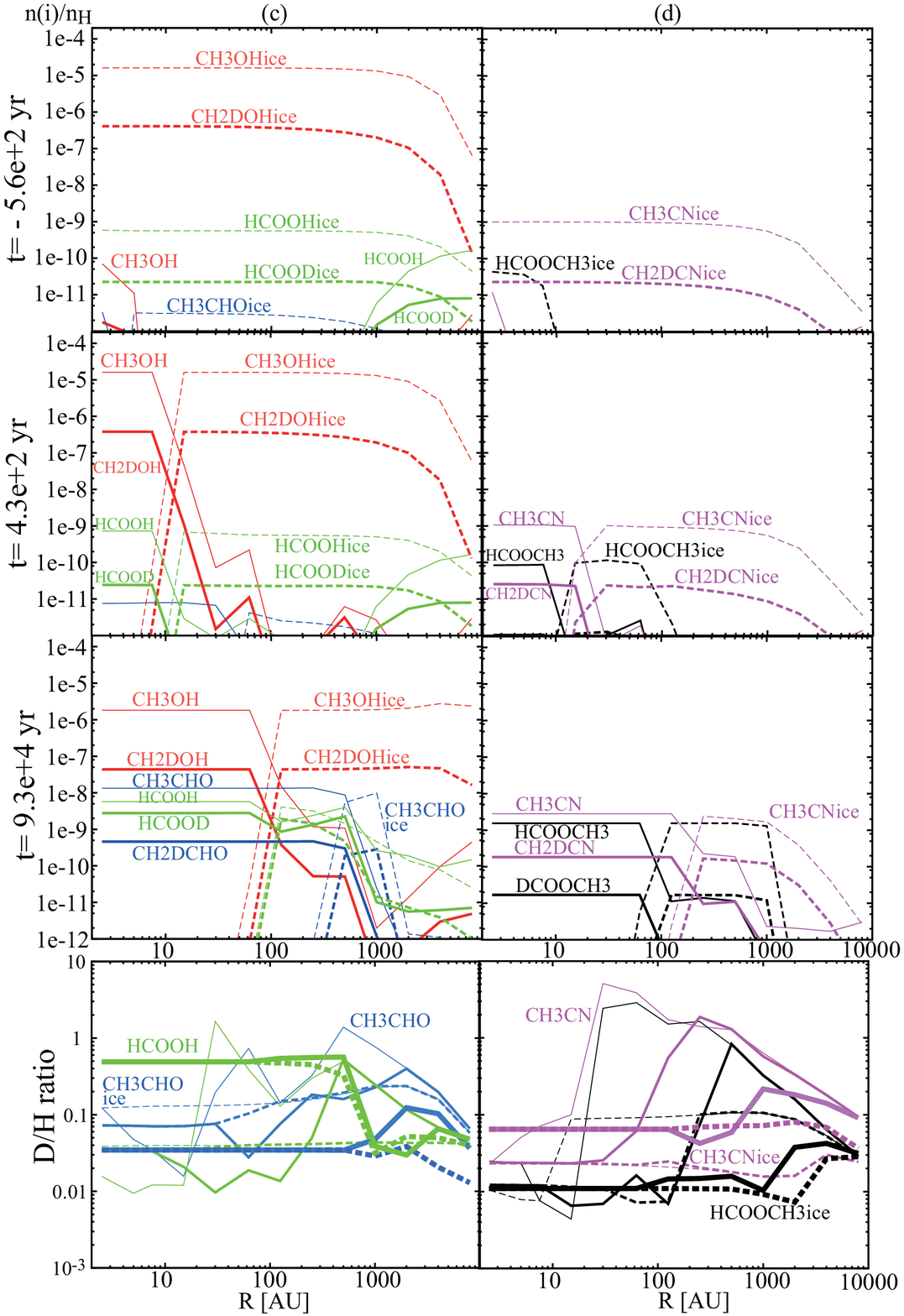}
\caption{cont.}
\end{figure}

\setcounter{figure}{2}
\begin{figure}
\epsscale{0.5}
\plotone{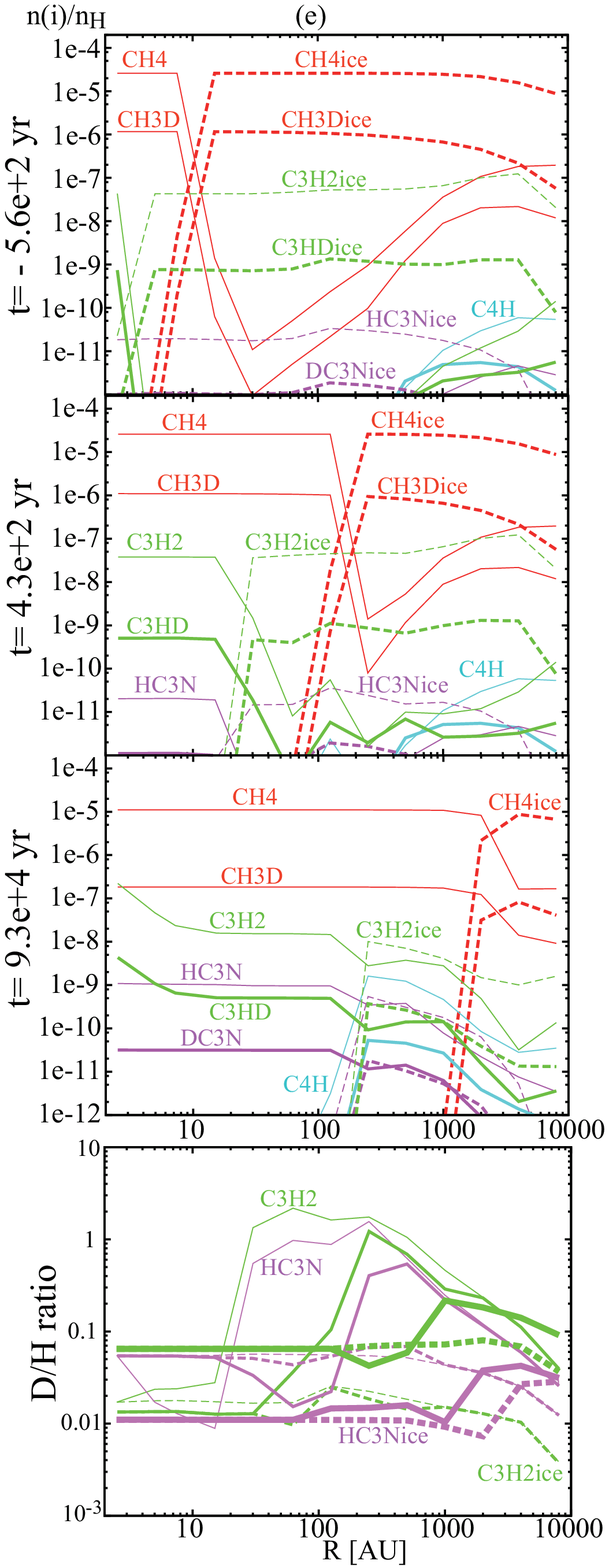}
\caption{cont.}
\end{figure}

\begin{figure}
\epsscale{1.0}
\plotone{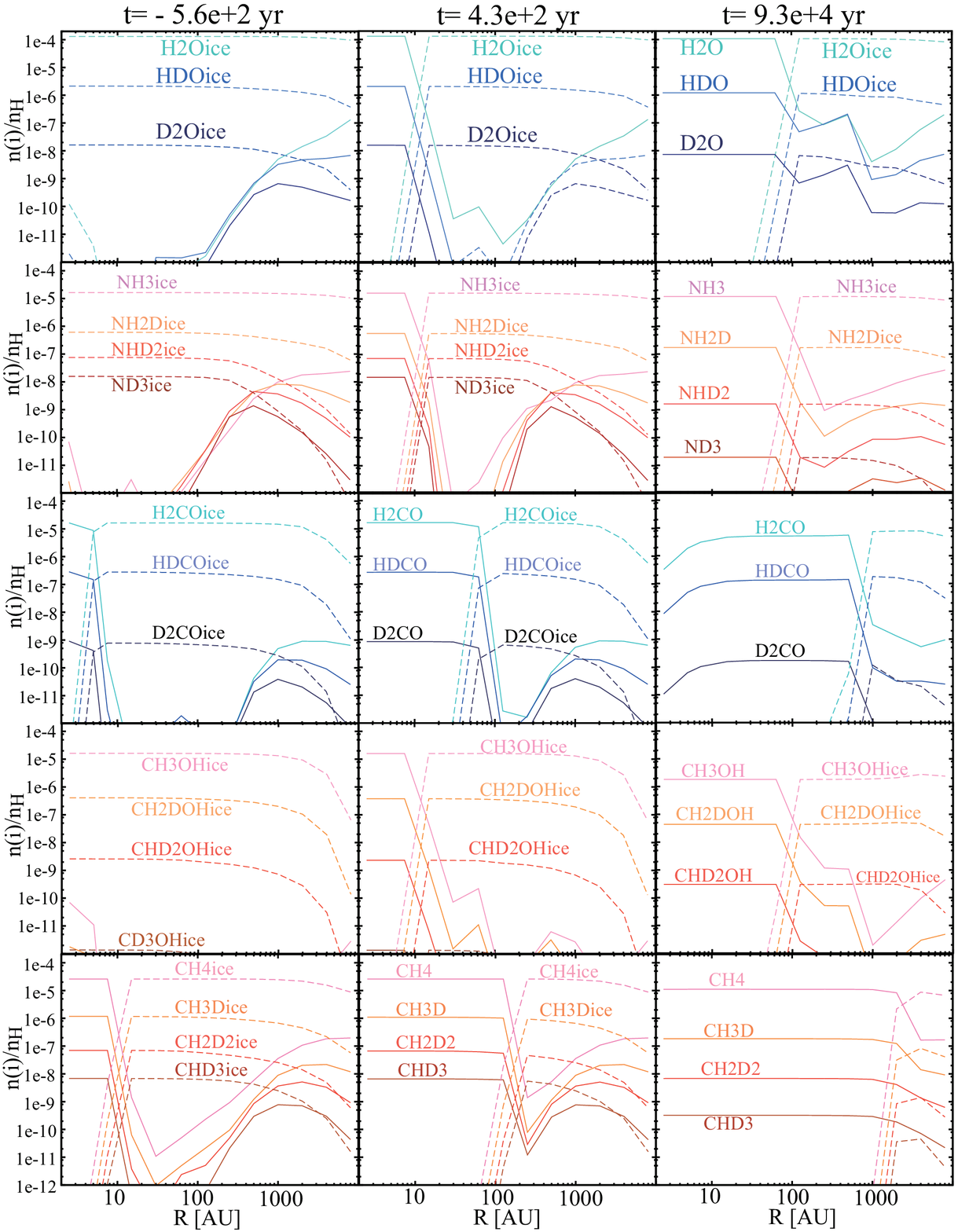}
\caption{Radial distribution of mono-, multi-deuterated species and their normal isotope
counterparts at $t=-5.6\times 10^2$ yr, $4.3\times 10^2$ yr, and $9.3\times 10^4$ yr.\label{dist_multiD}}
\end{figure}

\begin{figure}
\plotone{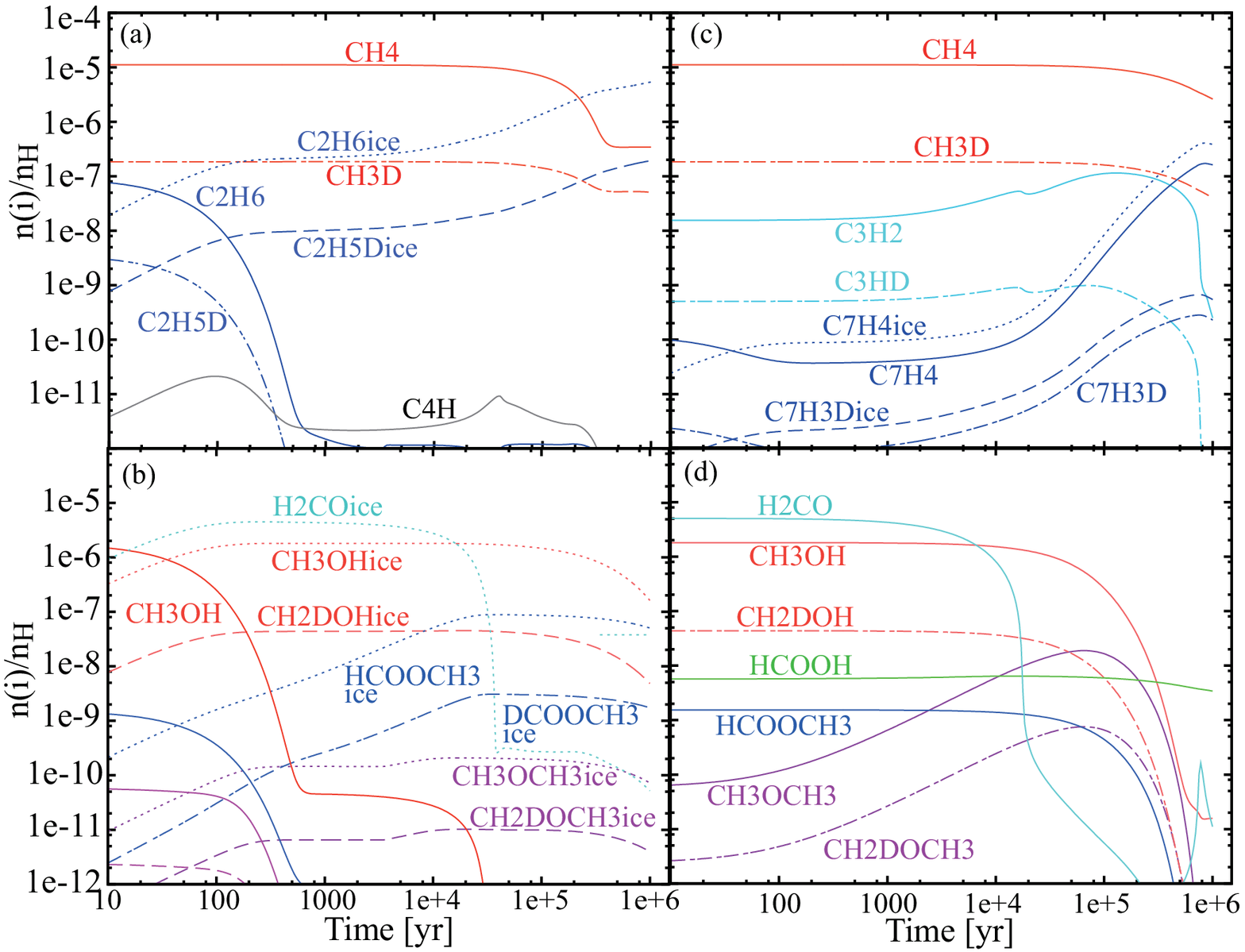}
\caption{Pseudo-time-dependent models with $n_{\rm H}=1.487\times 10^8$ cm$^{-3}$,
$T=40$ K (panels a, b) and $T=150$ K (panels c, d). The initial abundance is set by the 1D model
at $t=9.3\times 10^4$ yr and $r=30$ AU.
\label{disk}}
\end{figure}

\begin{figure}
\plotone{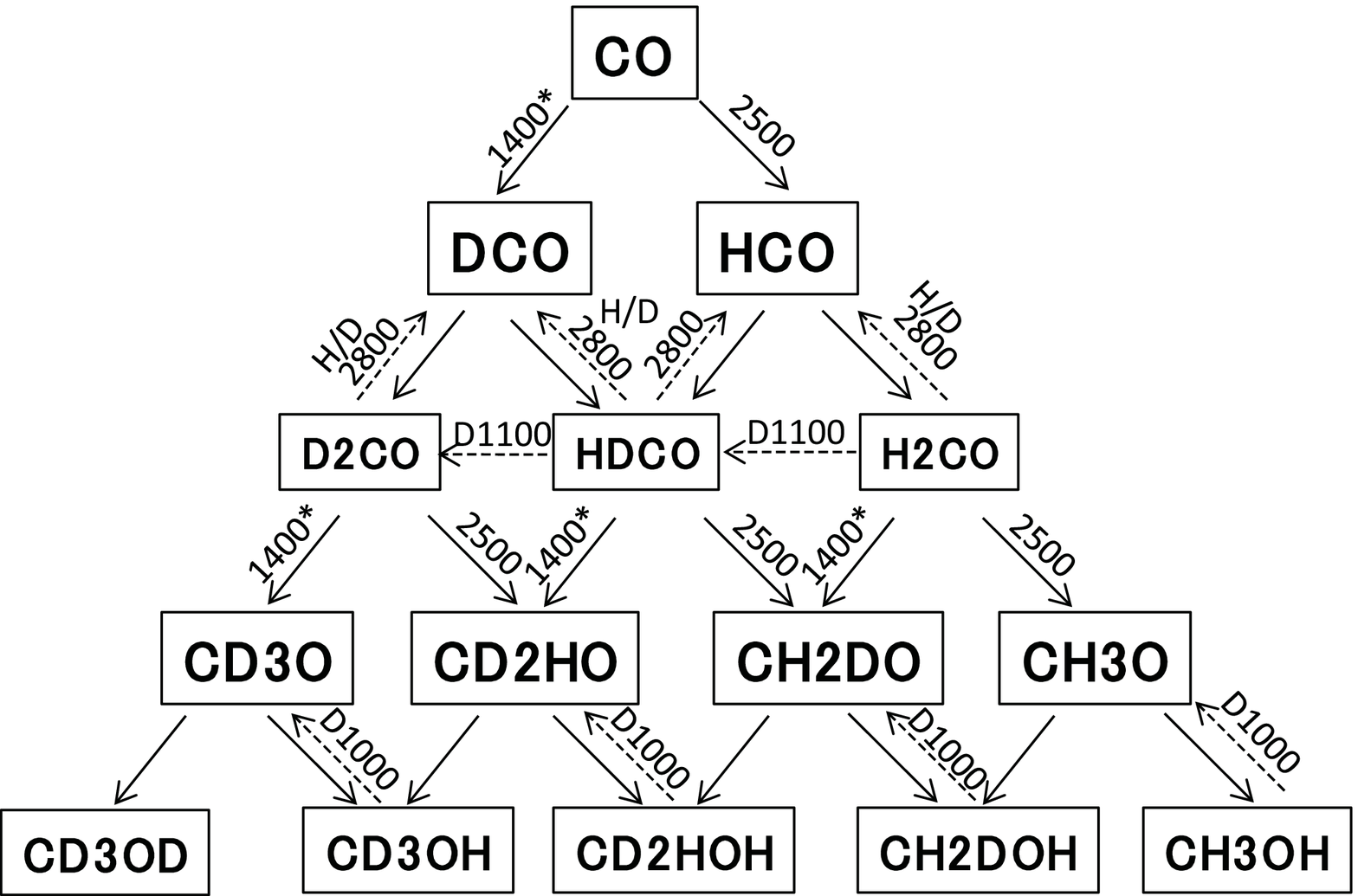}
\caption{Reaction network of H$_2$CO and CH$_3$OH formation on the grain surface. Solid arrows refer to the reactions
in our original model, while the dashed arrows are newly added reactions. The H or D and values represent the reactants
and activation barrier heights expediently assumed for the tunneling of our rectangular potential.
For D atom reactions with asterisks, we lowered the activation barrier
so that their rates are about 10 \% of the H atom reactions.
\label{Eact}}
\end{figure}

\begin{figure}
\epsscale{0.7}
\plotone{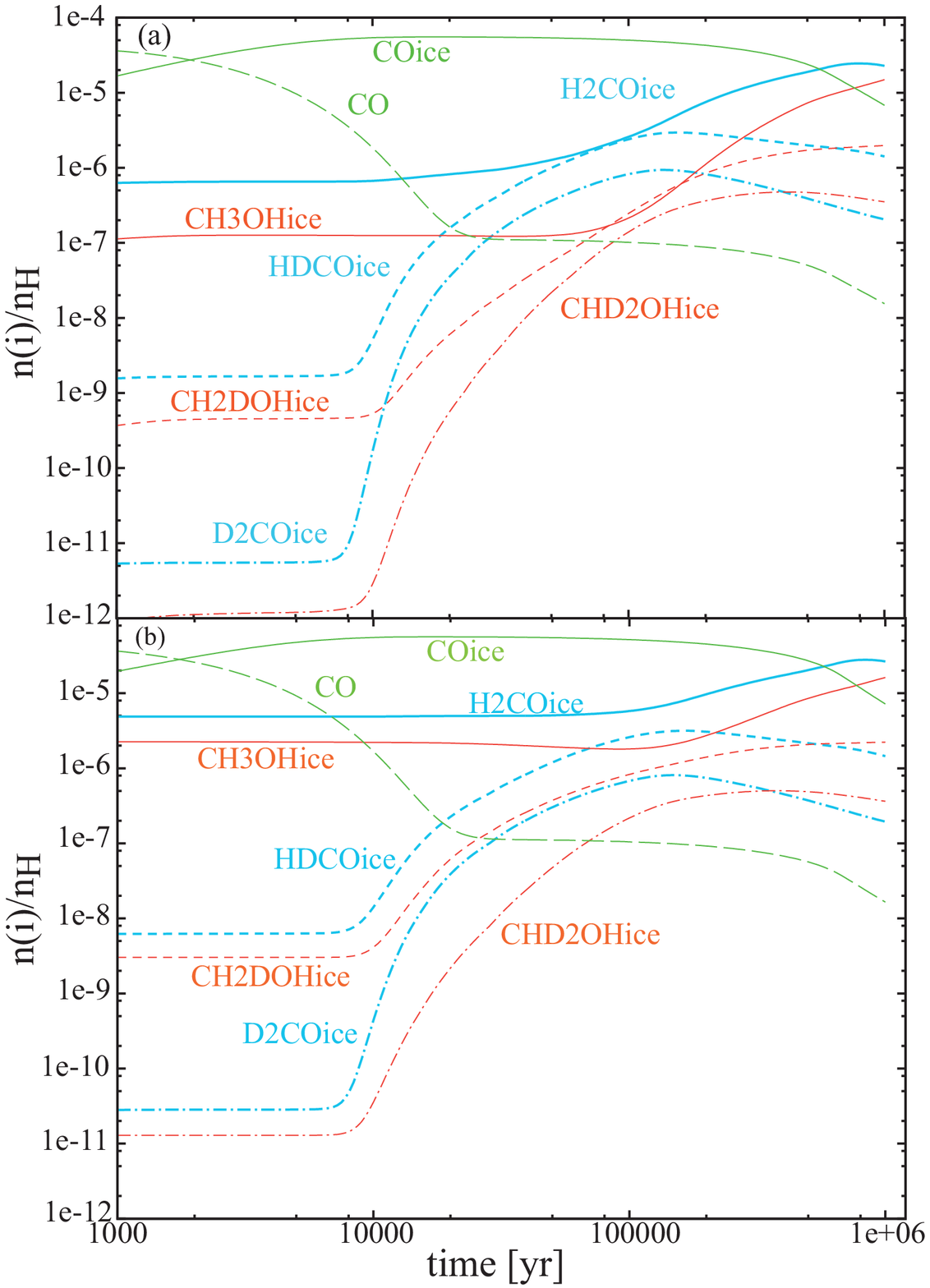}
\caption{Pseudo-time-dependent model at $n_{\rm H}=5\times 10^6$ cm$^{-3}$ and $T=10$ K with the substitution
and abstraction reactions. The initial abundance is determined by calculating the molecular evolution
at molecular cloud conditions for $8\times 10^4$ yr (a) and $2\times 10^5$ yr (b).
\label{cloud_CH3OH}}
\end{figure}
\end{document}